# Towards High-Efficiency Solar Cells: Insights into AsNCa$_3$ Antiperovskite as Active Layer

M. Irfan,[†,‡] B. D. Aparicio-Huacarpuma,[†,‡] C. M. de Oliveira Bastos,[¶] M. J. Piotrowski,[§] C. R. C. Rêgo,[∥] D. Guedes-Sobrinho,[⊥] R. Besse,[#] A. M. Almeida Silva,[@] Alexandre C. Dias,[*,#] and L. A. Ribeiro, Jr[*,†,‡]

[†]*Institute of Physics, University of Brasília, 70919-970, Brasília, DF, Brazil.*
[‡]*Computational Materials Laboratory, LCCMat, Institute of Physics, University of Brasília, 70919-970, Brasília, DF, Brazil.*
[¶]*Institute of Physics and International Center of Physics, University of Brasília, Brasília 70919-970, DF, Brazil*
[§]*Department of Physics, Federal University of Pelotas, PO Box 354, 96010-900, Pelotas, RS, Brazil.*
[∥]*Karlsruhe Institute of Technology (KIT), Institute of Nanotechnology, Hermann-von-Helmholtz-Platz, 76344, Eggenstein-Leopoldshafen, BW, Germany.*
[⊥]*Quantum Chemistry and Materials Thermodynamics group, Q$^2$M, Chemistry Department, Federal University of Paraná, CEP 81531 − 980, Curitiba, Brazil*
[#]*Institute of Physics and International Center of Physics, University of Brasília, 70919-970, Brasília, DF, Brazil.*
[@]*University of Brasília, College of Technology, Department of Mechanical Engineering, 70910-900, Brasília, Federal District, Brazil.*

E-mail: alexandre.dias@unb.br; ribeirojr@unb.br

## Abstract

Advances in photovoltaic technology are a viable route to contribute to cleaner and more sustainable energy solutions, placing perovskite-based materials among the best candidates for solar energy conversion. However, some challenges must be addressed to enhance their performance and stability. Herein, we report an investigation of the AsNCa$_3$ antiperovskite system for its potential in photovoltaic devices. We consider eight distinct crystalline phases, their structural parameters, dynamical stability, and electronic and optical properties. Furthermore, we consider each structural phase's contributions to solar harvesting efficiency by calculating the power conversion efficiency (PCE) using the spectroscopic limited maximum efficiency (SLME) formalism, which in this case reaches a maximum of 31.2 %. All dynamically stable phases exhibit a band gap around ∼ 1.3 eV, which lies within the optimal range for single-junction solar cells and yields PCE values comparable to the theoretical maximum PCE for silicon. These results place AsNCa$_3$ antiperovskites as promising candidates for high-efficiency photovoltaic applications. Notably, the PCE is only slightly changed by structural phase modification, suggesting that phase transitions induced by environmental conditions during device operation might not compromise device performance.

*Keywords*: AsNCa$_3$, Photovoltaics, Optoelectronic properties, Power conversion efficiency.

Solar energy promises to address the increasing global demand for electricity by harnessing photovoltaic (PV) technologies that convert direct sunlight into electric energy.[1,2] Of all the materials used in PV devices, halide perovskites are exceptional through their high absorption coefficient, long carrier diffusion length, and low-cost processing, among other benefits.[3,4] At the same time, however, regardless of the very high power conversion efficiencies that these materials, including organic-inorganic metal halide perovskites (MHPs),[5,6] are capable of, there exist several significant challenges. Among them, (*i*) thermal and moisture instability;[7,8] (*ii*) sensitivity to ultraviolet radiation;[9,10] (*iii*) chemical degradation by exposure to oxygen, water, and other species;[11–13] (*iv*) toxicity, as most of the efficient MHPs are lead-based (e.g., MAPbI$_3$, CsPbI$_3$, Cs$_2$PbI$_2$Cl$_2$, (BA)$_n$(MA)$_{n-1}$PbI$_4$);[8,10,14–17] (*v*) ion migration, wherein mobile species such as iodide and methylammonium migrate under electric fields or light exposure, causing changes in the structure and loss of efficiency.[18–20] Expanding the knowledge on chemically and structurally related materials becomes vital for overcoming these limitations.

The cubic phase of $\alpha$-CsPbI$_3$, one of the most representative all-inorganic MHPs, has been widely investigated, given its band gap close to 1.70 eV, which fits well within the optimal range for single-junction solar cells.[21–23] However, stabilizing this cubic phase at ambient conditions remains a challenge,[24–26] since at room temperature it spontaneously transforms into an orthorhombic phase ($\delta$-CsPbI$_3$, namely yellow phase) with a much larger band gap of 2.82 eV, i.e., the absorption spectrum is shifted towards the ultraviolet region.[27,28] In this context, beyond thermodynamic strategies such as pressure-induced stabilization,[29,30] manipulating the composition of ABX$_3$ represents a practical approach to correlate phase stability with desirable optoelectronic properties.[8,10,15,16]

Another perovskite-related family of materials, the antiperovskites, has also attracted notable interest in recent years due to their respective sets of novel properties, including superconductivity,[31], giant magnetoresistance,[32], and zero or negative thermal expansion.[33] Antiperovskites are characterized by the reversal of the A and X site occupancies of the classical perovskite structure, but still retaining the exact ideal cubic space group symmetry ($Pm\bar{3}m$).[1,34–36] Perhaps the most fascinating of these materials is AsNCa$_3$ due to its tunable band structures and likely topological character.[37] Additionally, its thermoelectric, piezoelectric[38], and charge transport behavior[39] are highly desirable for future energy applications and quantum devices.

Here, we investigate the photovoltaic (PV) potential of the antiperovskite AsNCa$_3$ using first-principles calculations. We systematically study its structural, electronic, and optical properties by advancing across eight different crystalline phases:[36] cubic, black, orthorhombic, yellow, hexagonal (2H and 4H), tetragonal, and a so-called supercubic phase,[36] which provides a more realistic description of cubic phase (a realistic distortion of the ideal cubic structure). Thus, the main idea is to understand how structural variations influence the material's light-harvesting performance and to identify promising phase configurations for photovoltaic applications.

We performed density functional theory (DFT) calculations using the Vienna *Ab Initio* Simulation Package (VASP),[40,41] solving the Kohn–Sham equations[42,43] via the projector augmented-wave (PAW) method.[44] The exchange-correlation energy was initially described using the Perdew–Burke–Ernzerhof (PBE) semilocal functional.[45] To compensate for the well-known underestimation of band gaps by local and semilocal functionals, which is attributable to self-interaction and derivative discontinuity errors,[46,47] we employed the HSE06 screened hybrid functional,[48,49] which typically reduces band gap discrepancies for common semiconductors. The structural optimization was conducted by relaxing both the stress tensor and atomic forces, with a plane-wave cutoff energy of 800 eV and a **k**-points density equivalent to 40 Å$^{-1}$. Convergence thresholds were set at $10^{-6}$ eV for total energy and 0.010 eV/Å for atomic forces. All subsequent calculations used a cutoff of 450 eV, while density of states (DOS) calculations employed a refined **k**-point density of 80 Å$^{-1}$.

Dynamical stability was assessed via phonon dispersion calculations at 0 K using the Phonopy package[50] in combination with the DFTB+ code,[51] based on the extended tight-binding method (xTB)[52] with GFN1-xTB parametrization.[53,54] For these calculations, the crystal structure was also optimized with DFTB+; the results are closer to the ones obtained with DFT, validating the approximation. This hybrid approach reproduces phonon frequencies at a lower computational cost than standard DFT phonon calculations. We used a **k**-point density of 40 Å$^{-1}$, with 3 × 3 × 3 supercells for the cubic phase and 2 × 2 × 2 supercells for all other phases to capture long-range interactions efficiently.

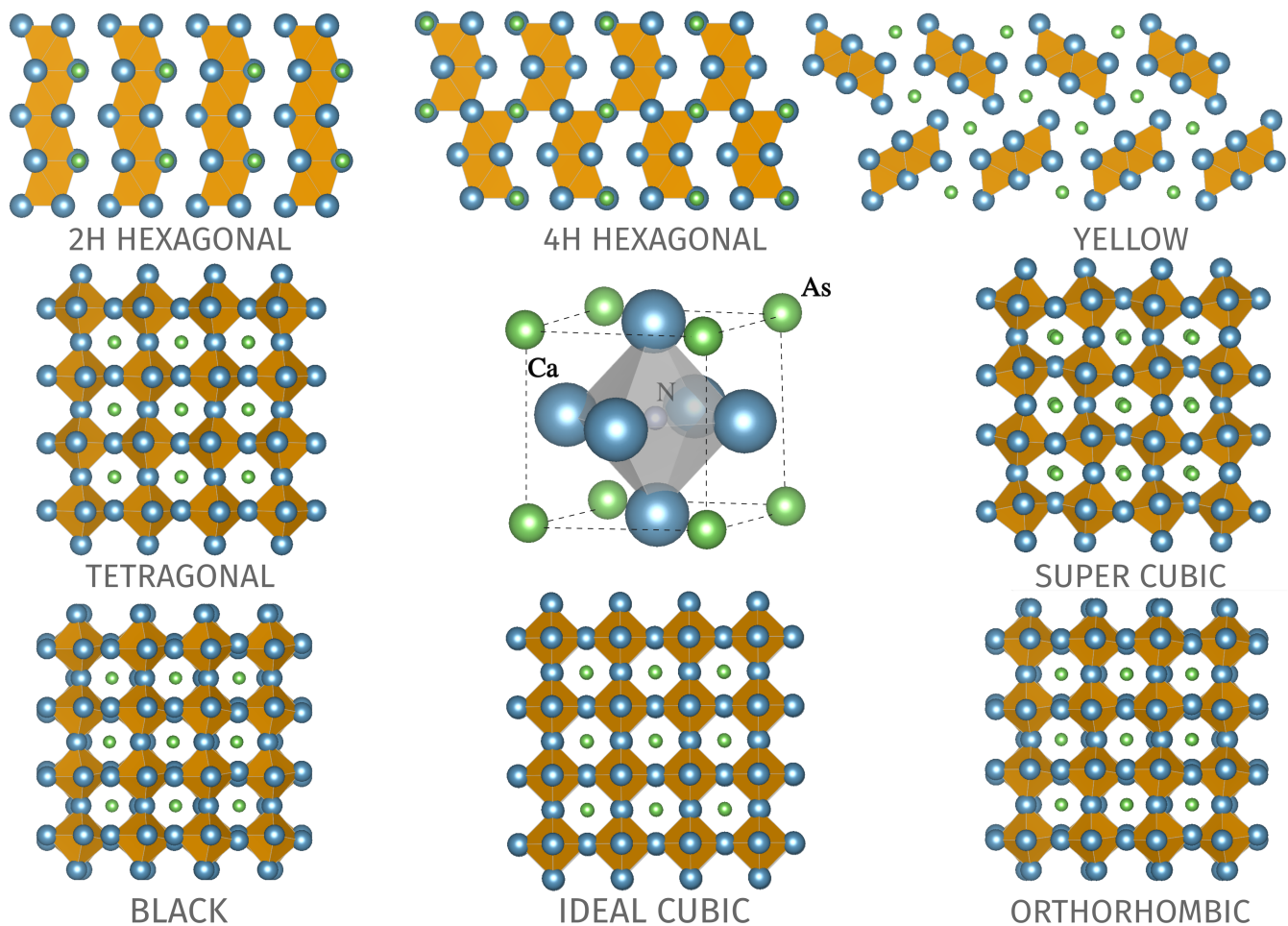

**Figure 1** Structural representations of the eight AsNCa$_3$ phases, highlighting different octahedral distortions.

Optical absorption spectra were computed within the independent-particle approximation (IPA) using WanTiBEXOS.[55] The calculations rely on a maximally localized Wannier function (MLWF) tight-binding Hamiltonian extracted from DFT–HSE06 calculations via the Wannier90 code,[56] with $s-$, $p-$ and $d-$orbital projections for As and Ca, and $p-$orbital projections for N. Full parameter details are provided in Section S10, Table S5, of the Supporting Information (SI) material. Finally, we evaluated the power conversion efficiency (PCE) using the spectroscopic limited maximum efficiency (SLME) framework,[57], assuming the AM1.5G solar spectrum[58] and device operation at 300 K.

Given the remarkable PCE presented in Fig. 5 for the AsNCa$_3$ antiperovskite. To ensure that our findings are fully reproducible[59,60], transferable[61,62], and compliant with the FAIR[63] data principles, we openly provide all key input files used throughout this work—including VASP, DFTB+, xTB, Wannier90, and WanTiBEXOS configurations—in a dedicated repository at https://github.com/ac-dias/AsNCa3-paper-raw-data. This archive also includes post-processing scripts and auxiliary data, enabling straightforward verification and extension of the present results.

The antiperovskite AsNCa$_3$ was examined in eight structural polymorphs, some of them previously identified by Dias et al.,[36] illustrated in Fig. 1. The number of formula units per cell (Z) varies by phase, as listed in Table 1. The assigned space groups are as follows: ideal cubic ($Pm\bar{3}m$); super cubic ($P1$, considered as a symmetry-lowered supercell of the cubic phase); tetragonal and orthorhombic (refined to appropriate space groups); black and yellow phases (both in $Pnma$); 2H hexagonal ($P6_3/m$); and 4H hexagonal ($Cmc2_1$). Complete atomic positions are provided in the SI. Table 1 reports equilibrium lattice parameters ($a_0$, $b_0$, and $c_0$), relative total energy per atom ($E_{rel}$), cohesive energies ($E_{coh}$), average N–Ca bond length (DAV), and effective coordination number (ECN). Note that ECN is calculated using a distance-weighted scheme (based on Hoppe's method)[64,65] to quantify the extent of octahedral distortion—values below 6 indicate deviations from ideal symmetry due to bond-length variation.

**Table 1** The equilibrium structural parameters of AsNCa$_3$ in eight different polymorphs, including the number of formula units per cell (Z), lattice constants ($a_0$, $b_0$, $c_0$), relative total energy per atom ($E_{rel}$), cohesive energy per atom ($E_{coh}$), effective coordination number (ECN), and average N-Ca bond length (DAV). Negative cohesive energies across all phases confirm favorable thermodynamic formation.

| Phase | Z | $a_0$ (Å) | $b_0$ (Å) | $c_0$ (Å) | $E_{rel}$ (eV/atom) | $E_{coh}$ (eV/atom) | ECN (N) | DAV (Å) |
|---|---|---|---|---|---|---|---|---|
| Ideal Cubic | 1 | 4.76 | 4.76 | 4.76 | 0.000 | −3.883 | 6.000 | 2.380 |
| Super Cubic | 8 | 9.53 | 9.53 | 9.53 | −0.006 | −3.890 | 6.000 | 2.408 |
| Tetragonal | 4 | 6.72 | 6.72 | 9.57 | −0.003 | −3.887 | 6.000 | 2.398 |
| Orthorhombic | 4 | 6.76 | 6.72 | 9.52 | −0.003 | −3.887 | 6.000 | 2.395 |
| Black | 4 | 6.73 | 6.74 | 9.56 | −0.006 | −3.890 | 6.000 | 2.410 |
| Yellow | 4 | 7.92 | 3.99 | 13.68 | 0.086 | −3.798 | 5.495 | 2.474 |
| 2H Hexagonal | 2 | 6.63 | 6.63 | 5.88 | 0.089 | −3.795 | 6.000 | 2.325 |
| 4H Hexagonal | 4 | 6.71 | 6.71 | 11.30 | 0.035 | −3.848 | 5.996 | 2.352 |

High-symmetry structures, namely ideal cubic, super-cubic, tetragonal, orthorhombic, and black, exhibit short energy differences ($\leq 6\,\text{meV/atom}$) and nearly constant average N–Ca bond lengths (within $\approx 0.030\,\text{Å}$), compared to lower symmetry phases (yellow, 2H, and 4H hexagonal). These observations suggest that such phases may interconvert under environmental perturbations (phase transitions), including solar illumination (energy influx from solar radiation). The ECN of N, in the center of NCa$_6$ octahedra, reveals the octahedral distortion, where more distorted structures show lower ECN values than the ideal (6). Furthermore, negative cohesive energies across all polymorphs confirm exothermic formation and favorable structural stability.

Phonon dispersion calculations were carried out to evaluate further the structural stability of the AsNCa$_3$ phases. As shown in Fig. 2, the absence of imaginary frequencies across the Brillouin zone confirms the dynamical stability of all considered phases, except for the 2H hexagonal phase, which exhibits negative (imaginary) phonon modes, indicative of dynamic instability. Complete phonon dispersion relations for all phases are provided in the SI (Section S6). Based on these results, only dynamically stable phases were considered in the subsequent electronic and optical properties analyses.

The electronic properties were investigated through band structure and DOS calculations. Representative results for the super cubic and yellow phases are presented in Fig. 3, while data for the remaining phases are available in the SI. In all cases, the conduction band minimum predominantly comprises Ca atomic states, regardless of the structural phase. However, the nature of the valence band maximum (VBM) varies with symmetry. In the high-symmetry phases, ideal cubic, super cubic, tetragonal, and black, the VBM is primarily derived from N orbitals, as seen in the super cubic example (Fig. 3). In contrast, for lower-symmetry structures such as the yellow, 2H, and 4H hexagonal phases (represented by the yellow phase in Fig. 3), the VBM is dominated by As orbitals. This shift in orbital character at the VBM is attributed to the symmetry breaking caused by octahedral distortions, which modify the local crystal field and orbital hybridization.[36]

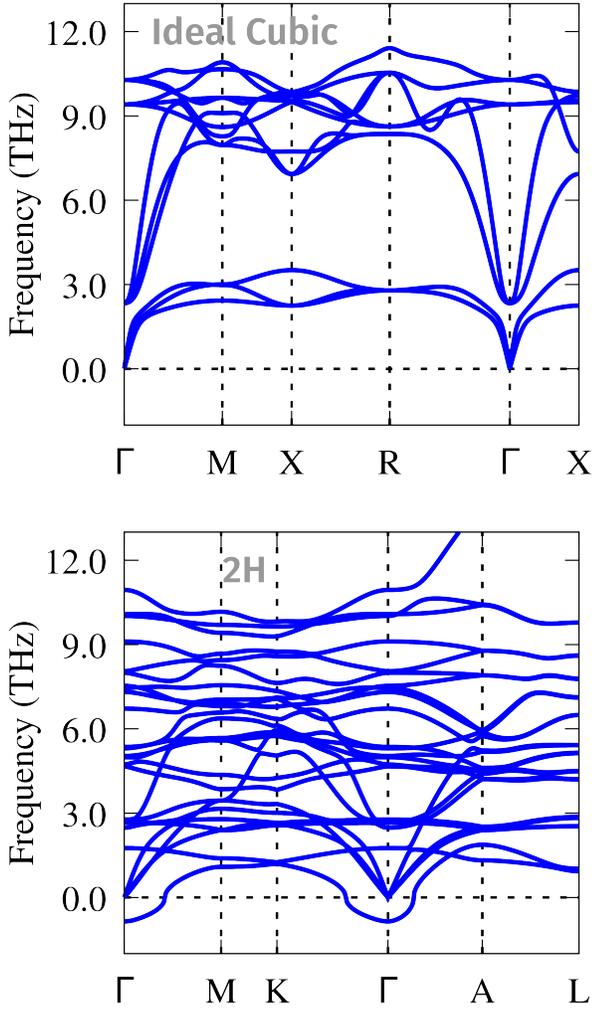

**Figure 2** Phonon dispersions for the ideal cubic (top panel) and 2H hexagonal (bottom panel) phases.

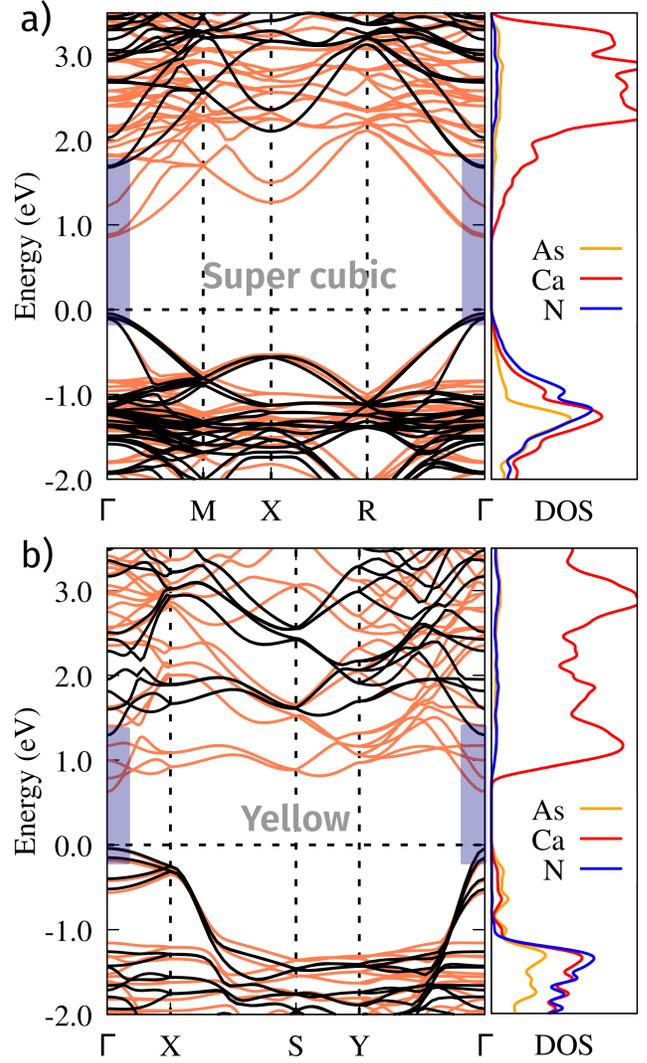

**Figure 3** Band structures and density of states (DOS) for $AsNCa_3$ in (a) Super cubic and (b) Yellow phases. The valence-band maximum has been shifted to 0 eV. Orange lines correspond to bands from PBE calculations and black lines to HSE06. Shaded regions highlight the topmost valence and bottommost conduction bands responsible for key optical transitions.

As well-established and reported in the literature,[36,66–68] the PBE functional underestimates band gaps, while HSE06 predictions align more closely with experimental observations. Therefore, the band gaps predicted using HSE06, as presented in Table 2, are considered more reliable for these materials. All dynamically stable phases exhibit a direct band gap at the $\Gamma$-point, as illustrated in Fig. 3. The computed gap values range from 1.33 eV to 1.72 eV, which falls within the optimal window (1.1–1.7 eV) for single-junction photovoltaic cells.[69] According to Shockley–Queisser theory, the maximum theoretical efficiency occurs at a band gap near 1.34 eV[69] (under AM1.5G illumination). Hence, based solely on band gap alignment, all examined phases demonstrate promising prospects for solar cell applications. This behavior contrasts with many halide perovskites, where phase transitions can significantly alter the band gap. In contrast, the similar band gap values across all $AsNCa_3$ phases suggest that efficiency should remain stable

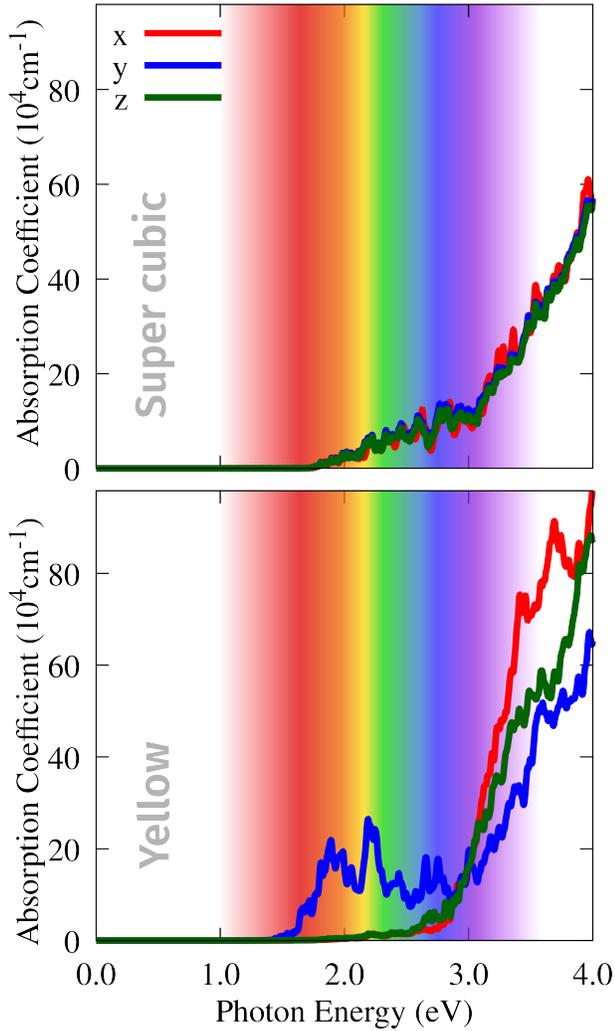

**Figure 4** Comparison of the absorption coefficient of AsNCa$_3$ in the Super cubic and Yellow phases. The plots show the linear absorption response under light polarized along the crystallographic directions $\hat{x}$ (red), $\hat{y}$ (blue), and $\hat{z}$ (green). The color-bar highlights the visible spectrum range, within which the material exhibits strong optical absorption.

even if phase conversions occur in response to environmental stimuli such as light or moisture.

To further explore the optoelectronic behavior of AsNCa$_3$, we investigated its linear optical response via the absorption coefficient (Fig. 4) for the super cubic (top) and yellow (bottom) phases. We found that most phases exhibit isotropic optical absorption, where the spectra for $\hat{x}$, $\hat{y}$, and $\hat{z}$ polarization (shown in red, blue, and green, respectively) are nearly identical. Conversely, the yellow phase displays optical anisotropy, reflecting its reduced symmetry, and absorption amplitudes vary notably across different polarizations in the visible range, as depicted in Fig. 4. Detailed values of absorption coefficients, refractive indices ($n_c$), and reflectivity ($R$) for all stable phases are provided in SI (Section S10).

**Table 2** Band gaps computed with PBE ($E_g^{\text{PBE}}$) and HSE06 ($E_g^{\text{HSE06}}$), alongside maximum theoretical power conversion efficiency (PCE$^{\text{SLME}}$) within the independent-particle approximation and at T = 300 K.

| Phase | $E_g^{\text{PBE}}$ (eV) | $E_g^{\text{HSE06}}$ (eV) | PCE$^{\text{SLME}}$ (%) |
|---|---|---|---|
| Ideal Cubic | 0.69 | 1.49 | 31.23 |
| Super Cubic | 0.90 | 1.72 | 27.50 |
| Tetragonal | 0.83 | 1.65 | 28.84 |
| Orthorhombic | 0.83 | 1.64 | 29.05 |
| Black | 0.90 | 1.72 | 30.24 |
| Yellow | 0.67 | 1.33 | 27.25 |
| 4H Hexagonal | 0.77 | 1.45 | 30.71 |

To evaluate the PV potential of AsNCa$_3$, we calculated the PCE using the SLME model as a function of absorber layer thickness (Fig. 5). Calculations were performed at 300 K within the IPA, considering film thicknesses up to 1 µm to minimize defect-mediated recombination. Table 2 summarizes the maximum SLME values. Among the polymorphs, the ideal cubic phase exhibits the highest PCE of 31.23%, while the yellow phase shows the lowest at 27.25%.

Despite the yellow phase band gap (1.33 eV) being very close to the Shockley–Queisser limit of 1.34 eV, it fails to reach the maximum SLME. This discrepancy arises from its optical anisotropy, an uneven absorption profile across polarizations, which reduces photon harvesting efficiency, particularly at lower energies. Nevertheless, increasing the film thickness above 1 µm can partly mitigate this effect.

Importantly, all stable phases deliver SLME values near or above the theoretical limit for silicon single-junction cells ($\approx 29\%$).[70] The consistency in high efficiencies across phases suggests photovoltaic performance will remain largely unaffected by environment-driven phase transitions, such as those induced by light or moisture, even considering the lowest values in Table 2.

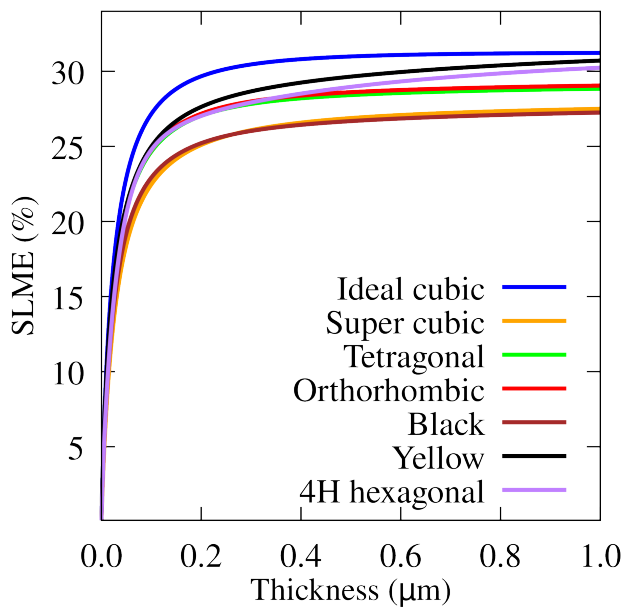

**Figure 5** PCE$^{SLME}$ *versus* **AsNCa$_3$ antiperovskite thickness for all stable phases.**

By analyzing the optoelectronic properties of AsNCa$_3$ antiperovskite polymorphs, we demonstrate their strong potential for single-junction solar-cell applications, achieving estimated SLME efficiencies near the theoretical maximum. Our findings also reveal that unlike many halide perovskites, whose band gaps and conversion efficiencies are significantly degraded by phase transitions, AsNCa$_3$ maintains consistent photovoltaic performance across all stable structural forms, with efficiencies closely tracking those of silicon ($\approx 29\%$) and exceeding it in some cases. Nevertheless, the device operational stability of AsNCa$_3$ over the long term must be investigated, particularly under conditions relevant to device deployment (e.g., exposure to moisture, ultraviolet light, and device-level electrical bias). What was learned from halide perovskite solar cells is the value of tailored passivation schemes and robust encapsulation for removing degradation mechanisms associated with environmental stressors and ionic migration. The same strategies, including grain boundaries, interfaces, and moisture barriers, will likely be required to realize the full potential of antiperovskite-based photovoltaics. These challenges must be overcome to translate the encouraging intrinsic properties of AsNCa$_3$ into high-performance, stable solar-cell devices.

# References

(1) Rani, U.; Kamlesh, P. K.; Shukla, A.; Verma, A. S. Emerging potential antiperovskite materials ANX3 (A= P, As, Sb, Bi; X= Sr, Ca, Mg) for thermoelectric renewable energy generators. *Journal of Solid State Chemistry* **2021**, *300*, 122246, DOI: 10.1016/j.jssc.2021.122246.

(2) Ali, R.; Hou, G.-J.; Zhu, Z.-G.; Yan, Q.-B.; Zheng, Q.-R.; Su, G. Predicted Lead-Free Perovskites for Solar Cells. *Chemistry of Materials* **2018**, *30*, 718–728, DOI: 10.1021/acs.chemmater.7b04036.

(3) Dong, Q.; Fang, Y.; Shao, Y.; Mulligan, P.; Qiu, J.; Cao, L.; Huang, J. Electron-hole diffusion lengths > 175 $\mu$m in solution-grown CH$_3$NH$_3$PbI$_3$ single crystals. *Science* **2015**, *347*, 967–970, DOI: 10.1126/science.aaa5760.

(4) Yin, W.-J.; Yang, J.-H.; Kang, J.; Yan, Y.; Wei, S.-H. Halide perovskite materials for solar cells: a theoretical review. *Journal of Materials Chemistry A* **2015**, *3*, 8926–8942, DOI: 10.1039/C4TA05033A.

(5) Sajid, S.; Alzahmi, S.; Salem, I. B.; Tabet, N.; Haik, Y.; Obaidat, I. M. Desirable candidates for high-performance lead-free organic–inorganic halide perovskite solar cells. *Materials for Renewable and Sustainable Energy* **2024**, *13*, 133–153, DOI: 10.1007/s40243-024-00255-w.

(6) Fan, X. Advanced progress in metal halide perovskite solar cells: A review. *Materials Today Sustainability* **2023**, *24*, 100603, DOI: 10.1016/j.mtsust.2023.100603.

(7) Meng, Q.; Chen, Y.; Xiao, Y. Y.; Sun, J.; Zhang, X.; Han, C. B.; Gao, H.; Zhang, Y.; Yan, H. Effect of temperature on the performance of perovskite solar cells. *Journal of Materials Science: Materials in Electronics*

**2020**, *32*, 12784–12792, DOI: 10.1007/s10854-020-03029-y.

(8) Araujo, L. O. d.; Rêgo, C. R.; Wenzel, W.; Bastos, C. M. d. O.; Piotrowski, M. J.; Dias, A. C.; Guedes-Sobrinho, D. Thermodynamic modeling and electronic properties of CsPb1−xSnxI3 as a polymorphic alloy. *Journal of Alloys and Compounds* **2024**, *992*, 174485, DOI: 10.1016/j.jallcom.2024.174485.

(9) Nie, W.; Blancon, J.-C.; Neukirch, A. J.; Appavoo, K.; Tsai, H.; Chhowalla, M.; Alam, M. A.; Sfeir, M. Y.; Katan, C.; Even, J.; Tretiak, S.; Crochet, J. J.; Gupta, G.; Mohite, A. D. Light-activated photocurrent degradation and self-healing in perovskite solar cells. *Nature Communications* **2016**, *7*, 11574, DOI: 10.1038/ncomms11574.

(10) Octavio de Araujo, L.; Rêgo, C. R. C.; Wenzel, W.; Sabino, F. P.; Guedes-Sobrinho, D. Impact of the Polymorphism and Relativistic Effects on the Electronic Properties of Inorganic Metal Halide Perovskites. *The Journal of Physical Chemistry C* **2022**, *126*, 2131–2140, DOI: 10.1021/acs.jpcc.1c08923.

(11) Mesquita, I.; Andrade, L.; Mendes, A. Effect of relative humidity during the preparation of perovskite solar cells: Performance and stability. *Solar Energy* **2020**, *199*, 474–483, DOI: 10.1016/j.solener.2020.02.052.

(12) Idigoras, J.; Aparicio, F. J.; Contreras-Bernal, L.; Ramos-Terrón, S.; Alcaire, M.; Sánchez-Valencia, J. R.; Borras, A.; Barranco, A.; Anta, J. A. Enhancing Moisture and Water Resistance in Perovskite Solar Cells by Encapsulation with Ultrathin Plasma Polymers. *ACS Applied Materials & Interfaces* **2018**, *10*, 11587–11594, DOI: 10.1021/acsami.7b17824, PMID: 29553253.

(13) Afre, R. A.; Pugliese, D. Perovskite Solar Cells: A Review of the Latest Advances in Materials, Fabrication Techniques, and Stability Enhancement Strategies. *Micromachines* **2024**, *15*, 192, DOI: 10.3390/mi15020192.

(14) Guedes-Sobrinho, D.; Neves Silveira, D.; de Araujo, L. O.; Favotto Dalmedico, J.; Wenzel, W.; Pramudya, Y.; Piotrowski, M. J.; Rêgo, C. R. C. Revealing the impact of organic spacers and cavity cations on quasi-2D perovskites via computational simulations. *Scientific Reports* **2023**, *13*, 4446, DOI: 10.1038/s41598-023-31220-8.

(15) Silveira, D. N.; Dias, A. C.; de Araujo, L. O.; Queiroz, M. A.; Dalmedico, J. F.; de Oliveira Bastos, C. M.; Rêgo, C. R.; Piotrowski, M. J.; Guedes-Sobrinho, D. Excitonic properties and solar harvesting performance of Cs2ZnY2X2 as quasi-2D mixed-halide perovskites. *Journal of Alloys and Compounds* **2024**, *1007*, 176434, DOI: 10.1016/j.jallcom.2024.176434.

(16) Dalmedico, J. F.; Silveira, D. N.; de Araujo, L. O.; Wenzel, W.; Rêgo, C. R. C.; Dias, A. C.; Guedes-Sobrinho, D.; Piotrowski, M. J. Tuning Electronic and Structural Properties of Lead-Free Metal Halide Perovskites: A Comparative Study of 2D Ruddlesden-Popper and 3D Compositions. *ChemPhysChem* **2024**, *25*, DOI: 10.1002/cphc.202400118.

(17) Rêgo, C. R. C.; Wenzel, W.; Piotrowski, M. J.; Dias, A. C.; Maciel de Oliveira Bastos, C.; de Araujo, L. O.; Guedes-Sobrinho, D. Digital workflow optimization of van der Waals methods for improved halide perovskite solar materials. *Digital Discovery* **2025**, *4*, 927–942, DOI: 10.1039/D4DD00312H.

(18) Zhang, X.; Chen, X.; Chen, Y.; Nadege Ouedraogo, N. A.; Li, J.; Bao, X.; Han, C. B.; Shirai, Y.; Zhang, Y.; Yan, H. Rapid degradation behavior of encapsulated perovskite solar cells under light, bias voltage or heat fields. *Nanoscale Advances* **2021**, *3*, 6128–6137, DOI: 10.1039/d1na00495f.

(19) Du, Y.; Wan, S.; Xie, M.; Xia, Y.; Yang, W.; Wei, Z.; Zhu, Y.; Hua, Y.; Jin, Z.; Hong, D.;

Tian, Y. Electric-Field-Induced Ion Migration Behavior in Methylammonium Lead Iodide Perovskite. *The Journal of Physical Chemistry Letters* **2021**, *12*, 7106–7112, DOI: 10.1021/acs.jpclett.1c01803.

(20) Bae, S.; Kim, S.; Lee, S.-W.; Cho, K. J.; Park, S.; Lee, S.; Kang, Y.; Lee, H.-S.; Kim, D. Electric-Field-Induced Degradation of Methylammonium Lead Iodide Perovskite Solar Cells. *The Journal of Physical Chemistry Letters* **2016**, *7*, 3091–3096, DOI: 10.1021/acs.jpclett.6b01176.

(21) Eperon, G. E.; Paternò, G. M.; Sutton, R. J.; Zampetti, A.; Haghighirad, A. A.; Cacialli, F. o.; Snaith, H. J. Inorganic caesium lead iodide perovskite solar cells. *J. Mater. Chem. A* **2015**, *3*, 19688–19695, DOI: 10.1039/C5TA06398A.

(22) Protesescu, L.; Yakunin, S.; Bodnarchuk, M. I.; Krieg, F.; Caputo, R.; Hendon, C. H.; Yang, R. X.; Walsh, A.; Kovalenko, M. V. Nanocrystals of Cesium Lead Halide Perovskites ($CsPbX_3$, X = Cl, Br, and I): Novel Optoelectronic Materials Showing Bright Emission with Wide Color Gamut. *Nano Letters* **2015**, *15*, 3692–3696, DOI: 10.1021/nl5048779.

(23) Waleed, A.; Tavakoli, M. M.; Gu, L.; Hussain, S.; Zhang, D.; Poddar, S.; Wang, Z.; Zhang, R.; Fan, Z. All Inorganic Cesium Lead Iodide Perovskite Nanowires with Stabilized Cubic Phase at Room Temperature and Nanowire Array-Based Photod etectors. *Nano Letters* **2017**, *17*, 4951–4957, DOI: 10.1021/acs.nanolett.7b02101, PMID: 28735542.

(24) Swarnkar, A.; Marshall, A. R.; Sanehira, E. M.; Chernomordik, B. D.; Moore, D. T.; tians, J. A. C.; Chakrabarti, T.; Luther, J. M. Quantum dot–induced phase stabilization of $α$-$CsPbI_3$ perovskite for high-efficiency photovoltaics. *Science* **2016**, *354*, 92–95, DOI: 10.1126/science.aag2700.

(25) Bhatt, H.; Saha, R.; Goswami, T.; C. K., S.; Justice Babu, K.; Kaur, G.; Shukla, A.; Patel, M. S.; Rondiya, S. R.; Dzade, N. Y.; Ghosh, H. N. Charge Transfer Modulation in the $α$-$CsPbI_3$/$WS_2$ Heterojunction via Band-Tailoring with Elemental Ni Doping. *ACS Photonics* **2024**, *11*, 5367–5379, DOI: 10.1021/acsphotonics.4c01774.

(26) Nigmetova, G.; Yelzhanova, Z.; Zhumadil, G.; Parkhomenko, H. P.; Tilegen, M.; Zhou, X. n.; Pavlenko, V.; Beisenbayev, A.; Aidarkhanov, D.; Jumabekov, A. N.; Kaikanov, M.; Pham, T. T.; B alanay, M. P.; Lim, C.-K.; Wang, Y.; Hu, H.; Ng, A. Controlling the Growth of $Cs_2PbX_4$ Nanostructures Enhances the Stability of Inorganic Cesium-Based Perovskite Solar Cells for Potentia l Low Earth Orbit Applications. *ACS Applied Materials & Interfaces* **2025**, *17*, 31575–31591, DOI: 10.1021/acsami.5c03064, PMID: 40377364.

(27) Wang, B.; Novendra, N.; Navrotsky, A. Energetics, Structures, and Phase Transitions of Cubic and Orthorhombic Cesium Lead Iodide ($CsPbI3$) Polymorphs. *Journal of the American Chemical Society* **2019**, *141*, 14501–14504, DOI: 10.1021/jacs.9b05924, PMID: 31487167.

(28) Lyu, H.; Su, H.; Lin, Z. Two-Stage Dynamic Transformation from $δ$- to $α$-$CsPbI3$. *The Journal of Physical Chemistry Letters* **2024**, *15*, 2228–2232, DOI: 10.1021/acs.jpclett.3c03244, PMID: 38373310.

(29) Ke, F.; Wang, C.; Jia, C.; Wolf, N. R.; Yan, J.; Niu, S.; Devereaux, T. P.; Karunadasa, H. I.; Mao, W. L.; Lin, Y. Preserving a robust $CsPbI3$ perovskite phase via pressure-directed octahedral tilt. *Nature Communications* **2021**, *12*, 461, DOI: 10.1038/s41467-020-20745-5.

(30) Ke, F.; Yan, J.; Niu, S.; Wen, J.; Yin, K.; Yang, H.; Wolf, N. R.; Tzeng, Y.-K.; Karunadasa, H. I.; Lee, Y. S.; Mao, W. L.; Lin, Y. Cesium-mediated electron redistribution and electron-electron interaction in high-pressure metallic $CsPbI3$.

*Nature Communications* **2022**, *13*, 7067, DOI: 10.1038/s41467-022-34786-5.

(31) He, T.; Huang, Q.; Ramirez, A. P.; Wang, Y.; Regan, K. A.; Rogado, N.; Hayward, M. A.; Haas, M. K.; Slusky, J. S.; Inumara, K.; Zandbergen, H. W.; Ong, N. P.; Cava, R. J. Superconductivity in the non-oxide perovskite MgCNi3. *Nature* **2001**, *411*, 54–56, DOI: 10.1038/35075014.

(32) Kim, W.; Chi, E.; Kim, J.; Choi, H.; Hur, N. Close correlation among lattice, spin, and charge in the manganese-based antiperovskite material. *Solid State Communications* **2001**, *119*, 507–510, DOI: 10.1016/s0038-1098(01)00279-4.

(33) Hamada, T.; Takenaka, K. Giant negative thermal expansion in antiperovskite manganese nitrides. *Journal of Applied Physics* **2011**, *109*, DOI: 10.1063/1.3540604.

(34) Hassan, M.; Arshad, I.; Mahmood, Q. Computational study of electronic, optical and thermoelectric properties of X3PbO (X = Ca, Sr, Ba) anti-perovskites. *Semiconductor Science and Technology* **2017**, *32*, 115002, DOI: 10.1088/1361-6641/aa8afe.

(35) Wang, Y.; Zhang, H.; Zhu, J.; Lü, X.; Li, S.; Zou, R.; Zhao, Y. Antiperovskites with Exceptional Functionalities. *Advanced Materials* **2019**, *32*, DOI: 10.1002/adma.201905007.

(36) Dias, A. C.; Lima, M. P.; Da Silva, J. L. F. Role of Structural Phases and Octahedra Distortions in the Optoelectronic and Excitonic Properties of CsGeX3 (X = Cl, Br, I) Perovskites. *The Journal of Physical Chemistry C* **2021**, *125*, 19142–19155, DOI: 10.1021/acs.jpcc.1c05245.

(37) Goh, W. F.; Pickett, W. E. Topological and thermoelectric properties of double antiperovskite pnictides. *Journal of Physics: Condensed Matter* **2020**, *32*, 345502, DOI: 10.1088/1361-648x/ab86f1.

(38) Ma, X.; Liu, C.; Materzanini, G.; Rignanese, G.-M.; Ren, W. Ferroelectricity, triple-well potential, and negative piezoelectricity in the antiperovskite $\gamma$-Ag$_3$SI. *Physical Review B* **2024**, *110*, DOI: 10.1103/physrevb.110.184109.

(39) Tong, P.; Sun, Y.; Zhao, B.; Zhu, X.; Song, W. Influence of carbon concentration on structural, magnetic and electrical transport properties for antiperovskite compounds AlCxMn3. *Solid State Communications* **2006**, *138*, 64–67, DOI: 10.1016/j.ssc.2006.02.009.

(40) Kresse, G.; Hafner, J. Ab initiomolecular dynamics for open-shell transition metals. *Physical Review B* **1993**, *48*, 13115–13118, DOI: 10.1103/physrevb.48.13115.

(41) Kresse, G.; Furthmüller, J. Efficient iterative schemes forab initiototal-energy calculations using a plane-wave basis set. *Physical Review B* **1996**, *54*, 11169–11186, DOI: 10.1103/physrevb.54.11169.

(42) Kohn, W.; Sham, L. J. Self-Consistent Equations Including Exchange and Correlation Effects. *Physical Review* **1965**, *140*, A1133–A1138, DOI: 10.1103/PhysRev.140.A1133.

(43) Hohenberg, P.; Kohn, W. Inhomogeneous electron gas. *Physical Review* **1964**, *136*, DOI: 10.1103/PHYSREV.136.B864.

(44) Kresse, G.; Joubert, D. From ultrasoft pseudopotentials to the projector augmented-wave method. *Physical Review B* **1999**, *59*, 1758–1775, DOI: 10.1103/PhysRevB.59.1758.

(45) Perdew, J. P.; Burke, K.; Ernzerhof, M. Generalized Gradient Approximation Made Simple. *Physical Review Letters* **1996**, *77*, 3865–3868, DOI: 10.1103/physrevlett.77.3865.

(46) Cohen, A. J.; Mori-Sánchez, P.; Yang, W. Fractional charge perspective on the band gap in density-functional theory. *Physical Review*

$B$ **2008**, *77*, DOI: 10.1103/physrevb.77.115123.

(47) Crowley, J. M.; Tahir-Kheli, J.; Goddard, W. A. Resolution of the Band Gap Prediction Problem for Materials Design. *The Journal of Physical Chemistry Letters* **2016**, *7*, 1198–1203, DOI: 10.1021/acs.jpclett.5b02870.

(48) Heyd, J.; Scuseria, G. E.; Ernzerhof, M. Hybrid functionals based on a screened Coulomb potential. *The Journal of Chemical Physics* **2003**, *118*, 8207–8215, DOI: 10.1063/1.1564060.

(49) Krukau, A. V.; Vydrov, O. A.; Izmaylov, A. F.; Scuseria, G. E. Influence of the exchange screening parameter on the performance of screened hybrid functionals. *The Journal of Chemical Physics* **2006**, *125*, DOI: 10.1063/1.2404663.

(50) Togo, A.; Tanaka, I. First principles phonon calculations in materials science. *Scripta Materialia* **2015**, *108*, 1–5, DOI: 10.1016/j.scriptamat.2015.07.021.

(51) Hourahine, B.; Aradi, B.; Blum, V.; Bonafé, F.; Buccheri, A.; Camacho, C.; Cevallos, C.; Deshaye, M. Y.; Dumitrică, T.; Dominguez, A.; Ehlert, S.; Elstner, M.; van der Heide, T.; Hermann, J.; Irle, S.; Kranz, J. J.; Köhler, C.; Kowalczyk, T.; Kubař, T.; Lee, I. S.; Lutsker, V.; Maurer, R. J.; Min, S. K.; Mitchell, I.; Negre, C.; Niehaus, T. A.; Niklasson, A. M. N.; Page, A. J.; Pecchia, A.; Penazzi, G.; Persson, M. P.; Řezáč, J.; Sánchez, C. G.; Sternberg, M.; Stöhr, M.; Stuckenberg, F.; Tkatchenko, A.; Yu, V. W.-z.; Frauenheim, T. DFTB+, a software package for efficient approximate density functional theory based atomistic simulations. *The Journal of Chemical Physics* **2020**, *152*, DOI: 10.1063/1.5143190.

(52) Bannwarth, C.; Caldeweyher, E.; Ehlert, S.; Hansen, A.; Pracht, P.; Seibert, J.; Spicher, S.; Grimme, S. Extended tight-binding quantum chemistry methods. *WIREs Computational Molecular Science* **2020**, *11*, DOI: 10.1002/wcms.1493.

(53) Grimme, S.; Bannwarth, C.; Shushkov, P. A Robust and Accurate Tight-Binding Quantum Chemical Method for Structures, Vibrational Frequencies, and Noncovalent Interactions of Large Molecular Systems Parametrized for All spd-Block Elements (Z = 1–86). *Journal of Chemical Theory and Computation* **2017**, *13*, 1989–2009, DOI: 10.1021/acs.jctc.7b00118.

(54) Vicent-Luna, J. M.; Apergi, S.; Tao, S. Efficient Computation of Structural and Electronic Properties of Halide Perovskites Using Density Functional Tight Binding: GFN1-xTB Method. *Journal of Chemical Information and Modeling* **2021**, *61*, 4415–4424, DOI: 10.1021/acs.jcim.1c00432.

(55) Dias, A. C.; Silveira, J. F.; Qu, F. WanTiBEXOS: A Wannier based Tight Binding code for electronic band structure, excitonic and optoelectronic properties of solids. *Computer Physics Communications* **2023**, *285*, 108636, DOI: 10.1016/j.cpc.2022.108636.

(56) Mostofi, A. A.; Yates, J. R.; Lee, Y.-S.; Souza, I.; Vanderbilt, D.; Marzari, N. wannier90: A tool for obtaining maximally-localised Wannier functions. *Computer Physics Communications* **2008**, *178*, 685–699, DOI: 10.1016/j.cpc.2007.11.016.

(57) Yu, L.; Zunger, A. Identification of Potential Photovoltaic Absorbers Based on First-Principles Spectroscopic Screening of Materials. *Physical Review Letters* **2012**, *108*, DOI: 10.1103/physrevlett.108.068701.

(58) ASTM-G173-03 Standard Tables for Reference Solar Spectral Irradiances: Direct Normal and Hemispherical on 37° Tilted Surface, ASTM International, West Conshohocken, PA (2012). 2012; https://doi.org/10.1520/g0173-03r20.

(59) Bekemeier, S.; Caldeira Rêgo, C. R.; Mai, H. L.; Saikia, U.; Waseda, O.; Apel, M.; Arendt, F.; Aschemann, A.; Bayerlein, B.; Courant, R.; Dziwis, G.; Fuchs, F.; Giese, U.; Junghanns, K.; Kamal, M.; Koschmieder, L.; Leineweber, S.; Luger, M.; Lukas, M.; Maas, J.; Mertens, J.; Mieller, B.; Overmeyer, L.; Pirch, N.; Reimann, J.; Schröck, S.; Schulze, P.; Schuster, J.; Seidel, A.; Shchyglo, O.; Sierka, M.; Silze, F.; Stier, S.; Tegeler, M.; Unger, J. F.; Weber, M.; Hickel, T.; Schaarschmidt, J. Advancing Digital Transformation in Material Science: The Role of Workflows Within the MaterialDigital Initiative. *Advanced Engineering Materials* **2025**, *27*, DOI: 10.1002/adem.202402149.

(60) Mostaghimi, M.; Rêgo, C. R. C.; Haldar, R.; Wöll, C.; Wenzel, W.; Kozlowska, M. Automated Virtual Design of Organic Semiconductors Based on Metal-Organic Frameworks. *Frontiers in Materials* **2022**, *9*, DOI: 10.3389/fmats.2022.840644.

(61) Rêgo, C. R. C.; Schaarschmidt, J.; Schlöder, T.; Penaloza-Amion, M.; Bag, S.; Neumann, T.; Strunk, T.; Wenzel, W. SimStack: An Intuitive Workflow Framework. *Frontiers in Materials* **2022**, *9*, DOI: 10.3389/fmats.2022.877597.

(62) Soleymanibrojeni, M.; Caldeira Rego, C. R.; Esmaeilpour, M.; Wenzel, W. An active learning approach to model solid-electrolyte interphase formation in Li-ion batteries. *Journal of Materials Chemistry A* **2024**, *12*, 2249–2266, DOI: 10.1039/d3ta06054c.

(63) Schaarschmidt, J.; Yuan, J.; Strunk, T.; Kondov, I.; Huber, S. P.; Pizzi, G.; Kahle, L.; Bölle, F. T.; Castelli, I. E.; Vegge, T.; Hanke, F.; Hickel, T.; Neugebauer, J.; Rêgo, C. R. C.; Wenzel, W. Workflow Engineering in Materials Design within the BATTERY 2030+ Project. *Advanced Energy Materials* **2021**, *12*, DOI: 10.1002/aenm.202102638.

(64) Hoppe, R. The Coordination Number − an "Inorganic Chameleon". *Angew. Chem. Int. Ed.* **1970**, *9*, 25–34, DOI: 10.1002/anie.197000251.

(65) Hoppe, R. Effective Coordination Numbers (ECoN) and Mean Active Fictive Ionic Radii (MEFIR). *Z. Kristallogr.* **1979**, *150*, 23–52, DOI: 10.1524/zkri.1979.150.1-4.23.

(66) Dias, A. C.; Bragança, H.; de Mendonça, J. P. A.; Da Silva, J. L. F. Excitonic Effects on Two-Dimensional Transition-Metal Dichalcogenide Monolayers: Impact on Solar Cell Efficiency. *ACS Applied Energy Materials* **2021**, *4*, 3265–3278, DOI: 10.1021/acsaem.0c03039.

(67) Bastos, C. M.; Besse, R.; Silva, J. L. D.; Sipahi, G. M. Ab initio investigation of structural stability and exfoliation energies in transition metal dichalcogenides based on Ti-, V-, and Mo-group elements. *Physical Review Materials* **2019**, *3*, DOI: 10.1103/PhysRevMaterials.3.044002.

(68) Bastos, C. M.; Sabino, F. P.; Sipahi, G. M.; Silva, J. L. D. A comprehensive study of g-factors, elastic, structural and electronic properties of III-V semiconductors using hybrid-density functional theory. *Journal of Applied Physics* **2018**, *123*, DOI: 10.1063/1.5018325.

(69) Rühle, S. Tabulated values of the Shockley–Queisser limit for single junction solar cells. *Solar Energy* **2016**, *130*, 139–147, DOI: 10.1016/j.solener.2016.02.015.

(70) Tiedje, T.; Yablonovitch, E.; Cody, G.; Brooks, B. Limiting efficiency of silicon solar cells. *IEEE Transactions on Electron Devices* **1984**, *31*, 711–716, DOI: 10.1109/t-ed.1984.21594.

# Supporting Information:

# Towards High-Efficiency Solar Cells: Ab Initio Insights into AsNCa$_3$ Antiperovskite

Muhammad Irfan,[†,‡] Bill D. Aparicio-Huacarpuma,[†,‡] Carlos Maciel de Oliveira Bastos,[¶] M. J. Piotrowski,[§] C. R. C. Rêgo,[∥] D. Guedes-Sobrinho,[⊥] Rafael Besse,[#] Alexandre C. Dias,[#] and L. A. Ribeiro, Jr[*,†,‡]

[†]*Institute of Physics, University of Brasília, 70919-970, Brasília, DF, Brazil.*
[‡]*Computational Materials Laboratory, LCCMat, Institute of Physics, University of Brasília, 70919-970, Brasília, DF, Brazil.*
[¶]*Institute of Physics and International Center of Physics, University of Brasília, Brasília 70919-970, DF, Brazil*
[§]*Department of Physics, Federal University of Pelotas, PO Box 354, 96010-900, Pelotas, RS, Brazil.*
[∥]*Karlsruhe Institute of Technology (KIT), Institute of Nanotechnology, Hermann-von-Helmholtz-Platz, 76344, Eggenstein-Leopoldshafen, BW, Germany.*
[⊥]*Quantum Chemistry and Materials Thermodynamics group, $Q^2M$, Chemistry Department, Federal University of Paraná, CEP 81531 − 980, Curitiba, Brazil*
[#]*Institute of Physics and International Center of Physics, University of Brasília, 70919-970, Brasília, DF, Brazil.*

E-mail: ribeirojr@unb.br

# Contents

# S1 PAW Projectors: Computational and Technical Details

Table S1: Information of the selected PAW projectors, element, PAW-PBE projector name, POTCAR date of creation, number of valence electrons, $Z_{val}$ and maximum recommended cutoff energy (ENMAX)

| Element | POTCAR PAW_PBE | Date POTCAR | $Z_{val}$ | ENMAX (eV) |
|---|---|---|---|---|
| As | As | 09/22/2009 | 5 | 208.702 |
| N | N | 04/08/2002 | 5 | 400.000 |
| Ca | Ca_pv | 09/06/2000 | 8 | 119.559 |

# S2 Structural Properties

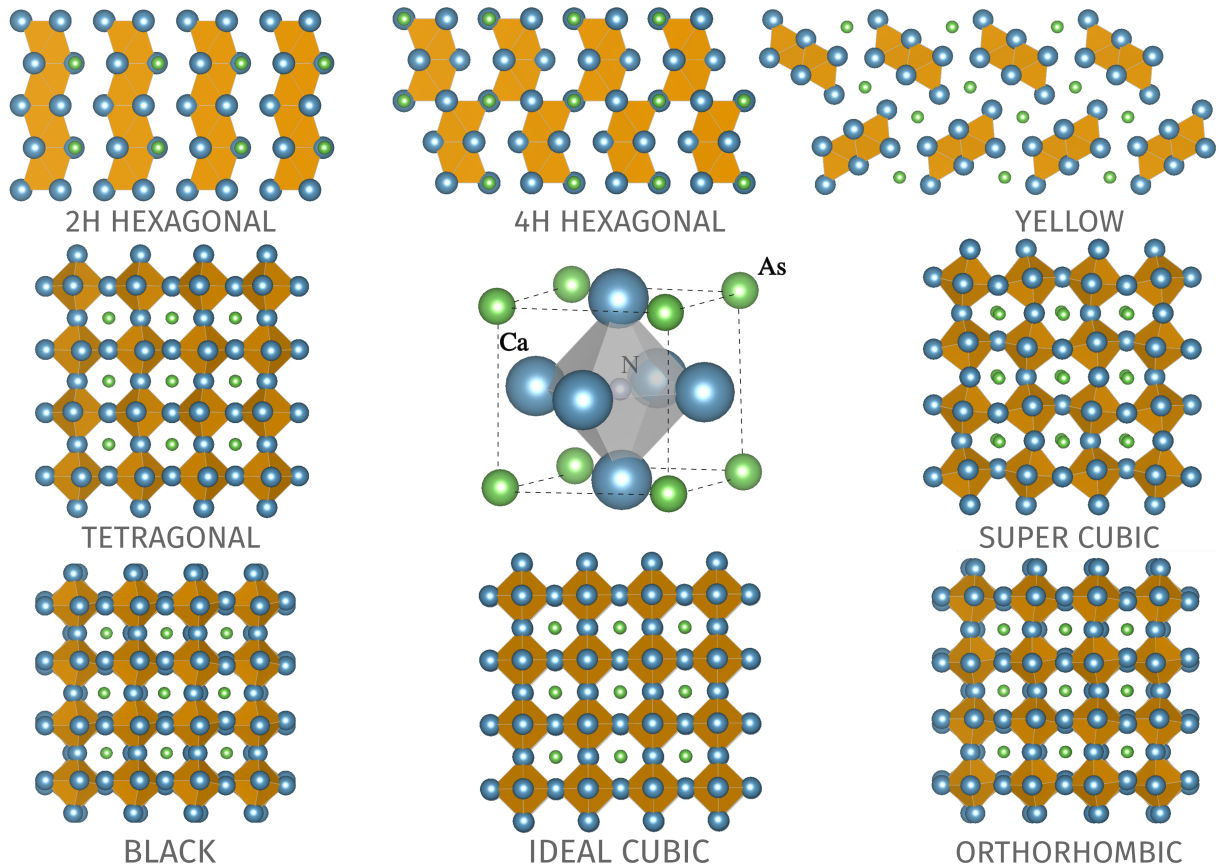

Figure S1: Representation of the AsNCa$_3$ anti-perovskites phases

Table S2: AsNCa$_3$ anti-perovskite octahedra effective coordination number (ECN) and average N-Ca bond length (DAV), for each structural phase.

| Phase | ECN (N) | DAV (Å) |
|---|---|---|
| Ideal Cubic | 6.000 | 2.380 |
| Super Cubic | 6.000 | 2.408 |
| Tetragonal | 6.000 | 2.398 |
| Orthorhombic | 6.000 | 2.395 |
| Black | 6.000 | 2.410 |
| Yellow | 5.495 | 2.474 |
| 2H Hexagonal | 6.000 | 2.325 |
| 4H Hexagonal | 5.996 | 2.352 |

# S3 Optimized POSCARs

## S3.1 Ideal Cubic

Listing 1: POSCAR Ideal Cubic Phase

```
AsNCa3
   1.00000000000000
     4.7605515385573289    0.0000000000000000    0.0000000000000000
     0.0000000000000000    4.7605515385573289    0.0000000000000000
     0.0000000000000000    0.0000000000000000    4.7605515385573289
   As   N   Ca
    1    1    3
Direct
  0.5000000000000000  0.5000000000000000  0.5000000000000000
  0.0000000000000000  0.0000000000000000  0.0000000000000000
  0.5000000000000000  0.0000000000000000  0.0000000000000000
  0.0000000000000000  0.5000000000000000  0.0000000000000000
  0.0000000000000000  0.0000000000000000  0.5000000000000000
```

## S3.2 Super Cubic

Listing 2: POSCAR Super Cubic Phase

```
AsNCa3-super_cubic
   1.00000000000000
```

```
    9.5347125281922107     0.0000000000000000     0.0000000000000000
    0.0000000000000000     9.5347125281922107     0.0000000000000000
    0.0000000000000000     0.0000000000000000     9.5347125281922107
  As   N   Ca
   8   8   24
Direct
  0.2582758922918913   0.2493357534072445   0.2664941639904157
  0.2582030742758690   0.2513680265707734   0.7685345713704521
  0.2581761157893112   0.7516030319356020   0.2687127393036093
  0.2583318004408781   0.7491787950768938   0.7660449759435082
  0.7582075129017483   0.2694799680486142   0.2510981540678046
  0.7581610338358900   0.2670689275112466   0.7484621868690624
  0.7581403671395037   0.7674704783980459   0.2486217600788265
  0.7582822648883365   0.7694364732564551   0.7505187255473871
  0.0082729471715979   0.0093802567494805   0.0085467310556169
  0.0083023380201013   0.0094252214584145   0.5085735615482889
  0.0082321475168072   0.5093847578330255   0.0085350689986257
  0.0082915818833627   0.5094504540899507   0.5085034711958585
  0.5082563806957552   0.0093008067709732   0.0086736560415304
  0.5081278868839263   0.0093697896784661   0.5087134915667875
  0.5082940929966639   0.5093625172952443   0.0086397893247252
  0.5081624323767571   0.5093968103309194   0.5086668932692007
  0.2582507808282628   0.9915066463390474   0.9844227909514629
  0.2582069125234554   0.0326336741599462   0.5260537386838067
  0.2582271620109893   0.5337524935958697   0.0258493710426606
  0.2581749753206140   0.4926039697530342   0.4845090443578215
  0.7582955759373604   0.0269281689231420   0.0326437059224887
  0.7582157749785878   0.9862853230135542   0.4913151730013894
  0.7582941209419474   0.4846184364925250   0.9911875550520364
  0.7582421889213080   0.5261818342905755   0.5324099004949545
  0.0297637481303568   0.2593544770907812   0.0407763459601398
  0.9878543146290042   0.2593596331214485   0.4765186546326987
  0.9864315012545291   0.7594057108057299   0.9765624422352062
  0.0292131092083352   0.7594416787816058   0.5406503082893366
  0.4867376145027862   0.2592453622704838   0.0402859094235666
```

```
38      0.5287742584666901   0.2593888932283122   0.4771064864335983
39      0.5304859104861848   0.7593079332905432   0.9770479419283404
40      0.4874908724375402   0.7593312117761286   0.5403046642266460
41      0.0295859492977826   0.9772920615932321   0.2584976392202165
42      0.9869442861850430   0.0416943366712488   0.7585036249711479
43      0.9868637004998675   0.5417043222027402   0.2584601172073491
44      0.0296681558732317   0.4774082512161257   0.7584462689067095
45      0.4866026286352039   0.9775649630309502   0.2587572770186028
46      0.5295562436720544   0.0409471675094935   0.7587582747014778
47      0.5296128329185024   0.5409433539169015   0.2586808004600201
48      0.4865169974974606   0.4774256462099630   0.7587297796185837
```

## S3.3 Orthorhombic

Listing 3: POSCAR Orthorhombic Phase

```
1   AsNCa3-orthorombic
2      1.00000000000000
3        6.7580408731249362    0.0000000000000000    0.0000000000000000
4        0.0000000000000000    6.7250557900332080    0.0000000000000000
5        0.0000000000000000    0.0000000000000000    9.5226315740136371
6      As   N   Ca
7       4   4   12
8   Direct
9    0.4992516487018506   0.9966080366152710   0.2491480333451435
10   0.0001617292444962   0.4968046029366207   0.2493717350014464
11   0.4999115393417242   0.9995627817176924   0.7505181773566676
12   0.9989851391085551   0.4991500462353926   0.7499187051255660
13   0.9997593829566469   0.9980743634088540   0.9996420166625342
14   0.4997556002364689   0.4981287700442465   0.9996797892148734
15   0.9995954875030932   0.9779642875003510   0.4997110399707481
16   0.4995814757682524   0.4980519713990716   0.4997273365384700
17   0.2380317949978519   0.2598342589798506   0.0220499777927827
18   0.7616817812996075   0.7364382937779581   0.9769587335436469
19   0.2618822631736322   0.7602204115118312   0.0197020533032202
```

```
20    0.7375710503261246   0.2360499065110560   0.9798728299942994
21    0.9581200007552937   0.9957919946267921   0.2497336755646558
22    0.5416111536438493   0.5014902869638718   0.2496826790762228
23    0.2670334692258578   0.2303879832963318   0.4773170784639404
24    0.7321386582834890   0.7656526797924101   0.5221895765833793
25    0.2315179034631640   0.7299223714183682   0.4803022152960708
26    0.7669254806839945   0.2657414351160625   0.5190930870731307
27    0.0416382472046379   0.0012577379078564   0.7496694237984585
28    0.4579040540182930   0.4953954956882782   0.7497412216121404
```

## S3.4 Tetragonal

Listing 4: POSCAR Tetragonal Phase

```
1  AsNCa3-tetragonal
2     1.00000000000000
3       6.7249684888785195   0.0000000000000000   0.0000000000000000
4       0.0000000000000000   6.7238375693452177   0.0000000000000000
5       0.0000000000000000   0.0000000000000000   9.5746436458272921
6     As    N    Ca
7      4    4    12
8  Direct
9    0.4981288796140717   0.9968334859638972   0.2502952208198863
10   0.9962658115121314   0.4986683437938950   0.2497886129036075
11   0.4961639983813058   0.9986887909783491   0.7493973334970647
12   0.9979476206072988   0.4969122619931596   0.7497475887073364
13   0.9970490937387311   0.9978022710776955   0.9997986126222642
14   0.4970199255027907   0.4977242485154889   0.9997806058810568
15   0.9970535452960121   0.9978052471324474   0.4998130157266161
16   0.4970373781701412   0.4977640440028068   0.4998158405469653
17   0.2812412312476624   0.2134831138807627   0.0083185214796870
18   0.7130935527366020   0.7814879695541208   0.9914421837633469
19   0.2132624449511340   0.7136948463784023   0.0000953499172027
20   0.7813585772907174   0.2825016893709886   0.9996935967779734
21   0.9871061027587018   0.9888231196396404   0.2495542009100902
```

```
22      0.5066203698096885   0.5066490194969617   0.2499467418058288
23      0.2140617818301394   0.2805139426398782   0.4914080288301648
24      0.7797361829786809   0.7149922640527535   0.5084451521710704
25      0.2797377709657454   0.7808223681240491   0.4996905649543564
26      0.7139997070151622   0.2144329549933133   0.5001978139796250
27      0.0066004068112093   0.0067060972915698   0.7499467653983913
28      0.4869829622075130   0.4884241468818402   0.7495638953234121
```

## S3.5 Black

Listing 5: POSCAR Black Phase

```
1  AsNCa3-black
2     1.00000000000000
3       6.7277114444613177   0.0000000000000000   0.0000000000000000
4       0.0000000000000000   6.7404496739393993   0.0000000000000000
5       0.0000000000000000   0.0000000000000000   9.5565743358256867
6     As   N   Ca
7      4   4   12
8  Direct
9   0.9975195241987080   0.4804377381065521   0.2500000000000000
10  0.0024804798012923   0.5195622618934479   0.7500000000000000
11  0.4975195241987080   0.0195622618934479   0.7500000000000000
12  0.5024804758012920   0.9804377381065521   0.2500000000000000
13  0.0000000000000000   0.0000000000000000   0.5000000000000000
14  0.5000000000000000   0.5000000000000000   0.5000000000000000
15  0.0000000000000000   0.0000000000000000   0.0000000000000000
16  0.5000000000000000   0.5000000000000000   0.0000000000000000
17  0.2150116318450301   0.2849939513859141   0.5209744096950146
18  0.7849883531549722   0.7150060786140813   0.4790255903049854
19  0.7150116468450278   0.2150060486140859   0.4790255903049854
20  0.2849883531549651   0.7849939213859187   0.5209744096950146
21  0.7849883531549722   0.7150060786140813   0.0209744096950146
22  0.2150116318450301   0.2849939513859141   0.9790255903049854
23  0.2849883531549651   0.7849939213859187   0.9790255903049854
```

```
24   0.7150116468450278  0.2150060486140859  0.0209744096950146
25   0.9597157284718634  0.9930682940982365  0.7500000000000000
26   0.0402842945281421  0.0069317219017648  0.2500000000000000
27   0.4597156984718609  0.5069317059017635  0.2500000000000000
28   0.5402842715281366  0.4930682640982340  0.7500000000000000
```

## S3.6  Yellow

Listing 6: POSCAR Yellow Phase

```
1   AsNCa3-yellow
2      1.000000000000000
3        7.9165560183520789    0.0000000000000000    0.0000000000000000
4        0.0000000000000000    3.9929219844398465    0.0000000000000000
5        0.0000000000000000    0.0000000000000000   13.6808644131219292
6      As   N   Ca
7       4   4   12
8   Direct
9    0.4141084253985738  0.2500000000000000  0.6671442305551238
10   0.5858915746014262  0.7500000000000000  0.3328557694448762
11   0.0858915746014262  0.7500000000000000  0.1671442305551238
12   0.9141084253985738  0.2500000000000000  0.8328557694448762
13   0.1465092874312361  0.2500000000000000  0.4387505775627716
14   0.8534906985687627  0.7500000000000000  0.5612493924372259
15   0.3534906985687627  0.7500000000000000  0.9387506075627741
16   0.6465093014312373  0.2500000000000000  0.0612494224372284
17   0.1770581413127559  0.2500000000000000  0.0048988555864469
18   0.8229418736872418  0.7500000000000000  0.9951011174135544
19   0.3229418736872418  0.7500000000000000  0.5048988825864456
20   0.6770581263127582  0.2500000000000000  0.4951011474135569
21   0.2922241618384618  0.2500000000000000  0.2926808090358648
22   0.7077758091615323  0.7500000000000000  0.7073192209641377
23   0.2077758381615382  0.7500000000000000  0.7926807790358623
24   0.7922241908384677  0.2500000000000000  0.2073191909641352
25   0.0236477160444011  0.2500000000000000  0.5997972916977261
```

```
0.9763522809555951    0.7500000000000000    0.4002027083022739
0.4763522809556022    0.7500000000000000    0.0997972916977261
0.5236477190444049    0.2500000000000000    0.9002027083022739
```

## S3.7   Hexagonal 2H

Listing 7: POSCAR Hexagonal 2H Phase

```
AsNCa3-2H
  1.00000000000000
     6.6294969521177869    0.0000000000000000    0.0000000000000000
    -3.3147468689072563    5.7413135790965075    0.0000000000000000
     0.0000000000000000    0.0000000000000000    5.8774985350970539
   As   N   Ca
    2   2   6
Direct
 0.6666666651404114    0.3333333863540702    0.2500000273156857
 0.3333333351701526    0.6666666283517984    0.7500000134463249
 0.0000000755941372    0.9999999610813859    0.5000000161388272
 0.9999999475910428    0.0000000527331707    0.9999999763237213
 0.3137628835006225    0.1568836794840749    0.7499999628419047
 0.6862370747804505    0.8431162837831749    0.2500000180868938
 0.8431162401193149    0.1568792404409578    0.7500000212267111
 0.1568837607627884    0.8431207777608378    0.2500000173755339
 0.8431208286625562    0.6862370904136910    0.7500000269216827
 0.1568791616785177    0.3137629045968424    0.2499999803227055
```

## S3.8   Hexagonal 4H

Listing 8: POSCAR Hexagonal 4H Phase

```
AsNCa3-4H
  1.00000000000000
     6.7083147767129034    0.0000000000000000    0.0000000000000000
    -3.3541549599152471    5.8095731282329712    0.0000000000000000
```

```
     0.0000000000000000     0.0000000000000000    11.3027754566063443
   As   N   Ca
    4    4   12
Direct
  0.9999999621681042  0.0000001168674046  0.5000000744719060
  0.0000001113783128  0.9999999423216508  0.0000001300835848
  0.3333330264231051  0.6666671010863610  0.7500001568988424
  0.6666671585464456  0.3333330348521173  0.2500000855702282
  0.3333337533942711  0.6666665106457828  0.3834079803784647
  0.6666665815026462  0.3333331963629433  0.6165917263280534
  0.6666664796554045  0.3333328451621398  0.8834083862403261
  0.3333327847709242  0.6666665296169896  0.1165917279868651
  0.4999994216651586  0.9999993277137094  0.4999999988831760
  0.0000007175603187  0.5000006387435221  0.5000000060913266
  0.4999999509930717  0.5000001476576585  0.5000004760297116
  0.5000006111542064  0.0000007676921143  0.9999999264184964
  0.9999992909048672  0.4999994268006063  0.9999999573967315
  0.5000000938624609  0.4999999514901887  0.0000004370697297
  0.1768364023707178  0.8231636263789923  0.2500004904843323
  0.8231635899975274  0.1768364442473001  0.7500004080479954
  0.1768369363685238  0.3536731430467697  0.2499994617217993
  0.8231634217342432  0.6463270591134176  0.7499995141687137
  0.6463269103398872  0.8231633990190304  0.2499996568397407
  0.3536727772097805  0.1768368111812748  0.7499993828899960
```

# S4  High Symmetry K-points

## S4.1  Ideal Cubic and Super Cubic

- G: 0.0 0.0 0.0

- M: 0.5 0.0 0.5

- X: 0.5 0.0 0.0

- R: 0.5 0.5 0.5

## S4.2 Hexagonal 2H and 4H

- G: 0.0 0.0 0.0

- M: 0.0 -0.5 0.0

- K: 0.3 -0.6 0.0

- A: 0.0 0.0 0.5

- L: 0.0 -0.5 0.5

- H: 0.3 -0.6 0.5

## S4.3 Black and Yellow Phase

- G: 0.0 0.0 0.0

- X: 0.5 0.0 0.0

- S: 0.5 0.5 0.0

- Y: 0.0 0.5 0.0

- Z: 0.0 0.0 0.5

- U: 0.5 0.0 0.5

- R: 0.5 0.5 0.5

- T: 0.0 0.5 0.5

## S4.4 Tetragonal

- G: 0.0 0.0 0.0

- M: 0.5 0.5 0.0

- X: 0.5 0.0 0.0
- Z: 0.0 0.0 0.5
- A: 0.5 0.0 0.5
- R: 0.5 0.5 0.5

## S4.5 Orthorhombic

- G: 0.0 0.0 0.0
- X: 0.5 0.0 0.0
- S: 0.5 0.5 0.0
- Y: 0.0 0.5 0.0
- Z: 0.0 0.0 0.5
- U: 0.5 0.0 0.5
- R: 0.5 0.5 0.5
- T: 0.0 0.5 0.5

## S5 Structural and Energetic Properties

Table S3: AsNCa$_3$ equilibrium crystal parameters include the number of atoms ($N_{atm}$) in unit cell, unit cell lattice parameters ($a_0$, $b_0$, $c_0$, $\alpha$, $\beta$, $\gamma$), relative total energy per atom ($E_{Rel}$) from ideal cubic phase, and cohesive energy per atom $E_{coh}$ for all investigated structural phases.

| Phase | $N_{atm}$ | $a_0$ (Å) | $b_0$ (Å) | $c_0$ (Å) | $\alpha$ (°) | $\beta$ (°) | $\gamma$ (°) | $E_{rel}$ (eV/atom) | $E_{coh}$ (eV/atom) |
|---|---|---|---|---|---|---|---|---|---|
| Ideal Cubic | 5 | 4.76 | 4.76 | 4.76 | 90 | 90 | 90 | 0.000 | −3.883 |
| Super Cubic | 40 | 9.53 | 9.53 | 9.53 | 90 | 90 | 90 | −0.006 | −3.890 |
| Tetragonal | 20 | 6.72 | 6.72 | 9.57 | 90 | 90 | 90 | −0.003 | −3.887 |
| Orthorhombic | 20 | 6.76 | 6.72 | 9.52 | 90 | 90 | 90 | −0.003 | −3.887 |
| Black | 20 | 6.73 | 6.74 | 9.56 | 90 | 90 | 90 | -0.006 | −3.890 |
| Yellow | 20 | 7.92 | 3.99 | 13.68 | 90 | 90 | 90 | 0.086 | −3.798 |
| 2H Hexagonal | 10 | 6.63 | 6.63 | 5.88 | 90 | 90 | 120 | 0.089 | −3.795 |
| 4H Hexagonal | 20 | 6.71 | 6.71 | 11.30 | 90 | 90 | 120 | 0.035 | −3.848 |

## S6 Phonon Dispersion

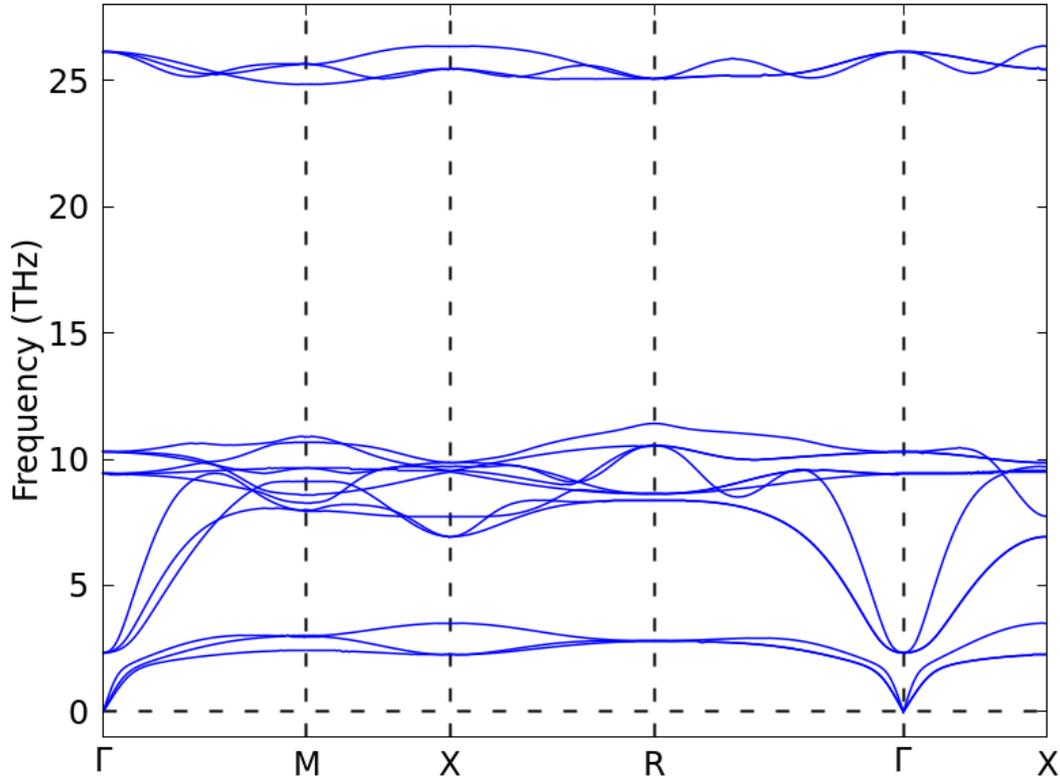

Figure S2: Phonon dispersion for AsNCa$_3$ antiperovskite in Ideal Cubic Phase.

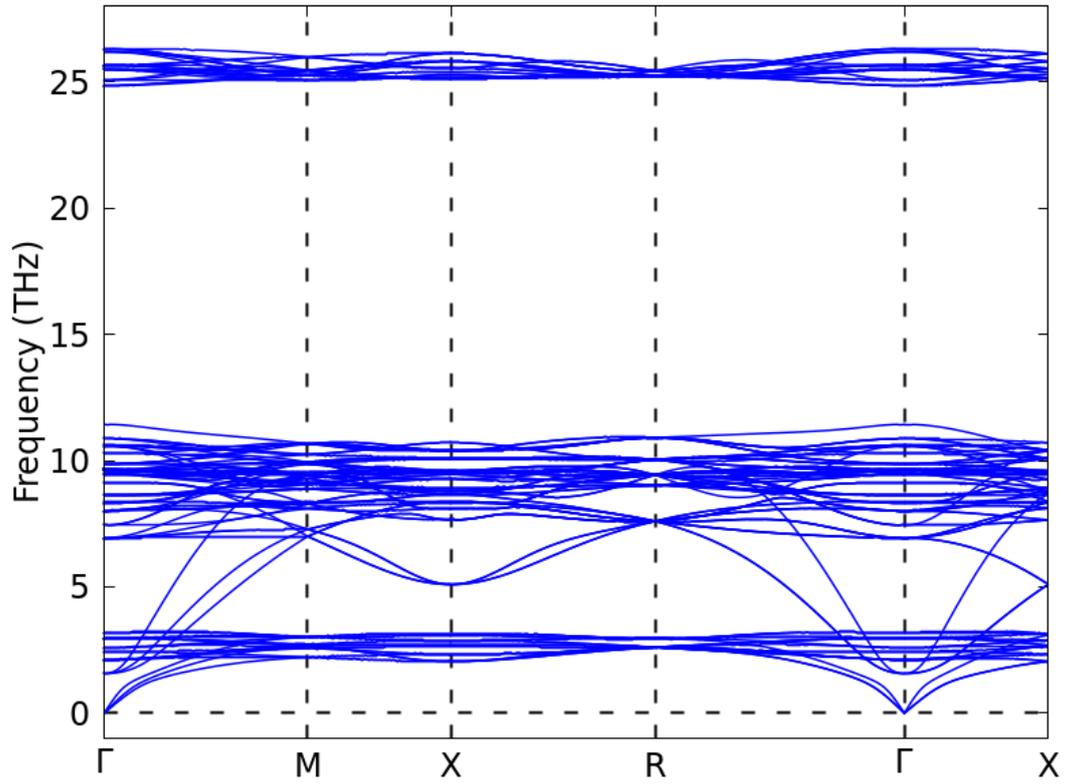

Figure S3: Phonon dispersion for AsNCa$_3$ antiperovskite in Super Cubic Phase.

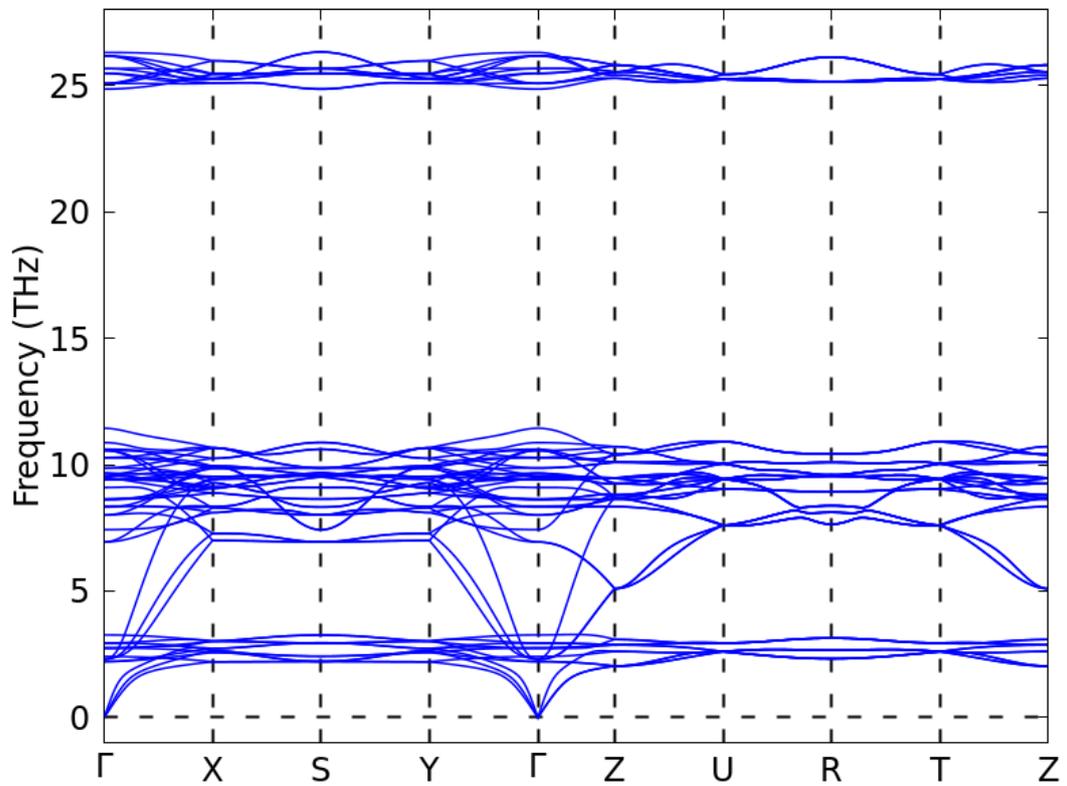

Figure S4: Phonon dispersion for AsNCa$_3$ antiperovskite in Orthorhombic Phase.

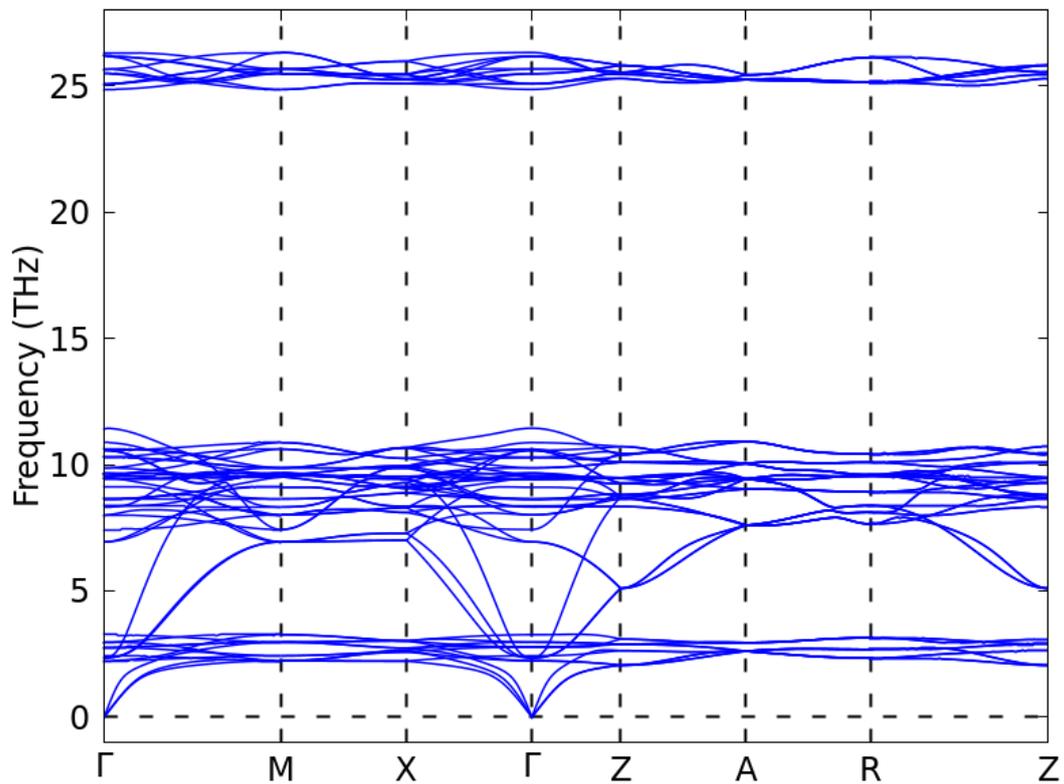

Figure S5: Phonon dispersion for AsNCa$_3$ antiperovskite in Tetragonal Phase.

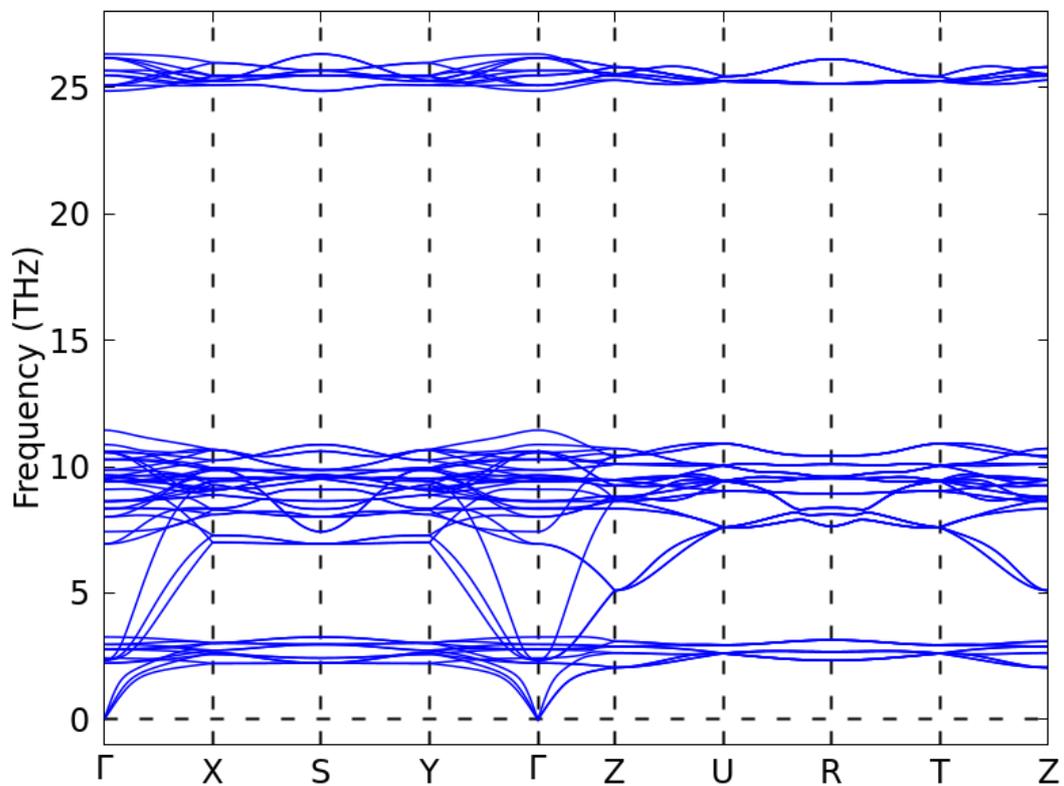

Figure S6: Phonon dispersion for AsNCa$_3$ antiperovskite in Black Phase.

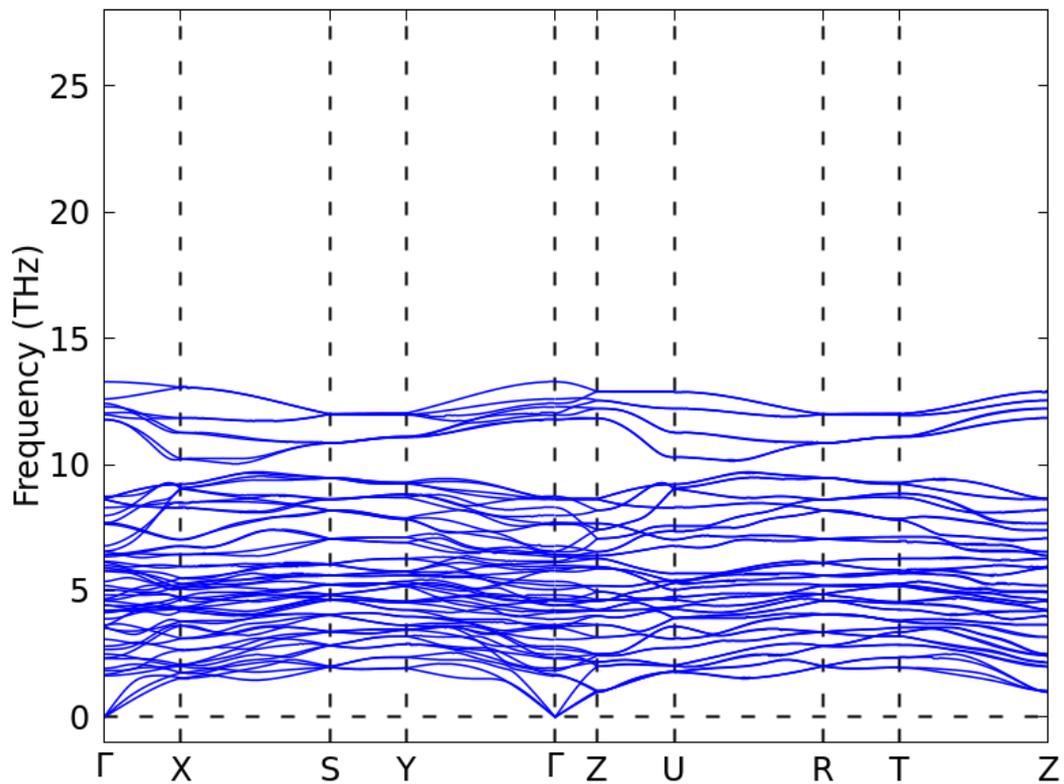

Figure S7: Phonon dispersion for AsNCa$_3$ antiperovskite in Yellow Phase.

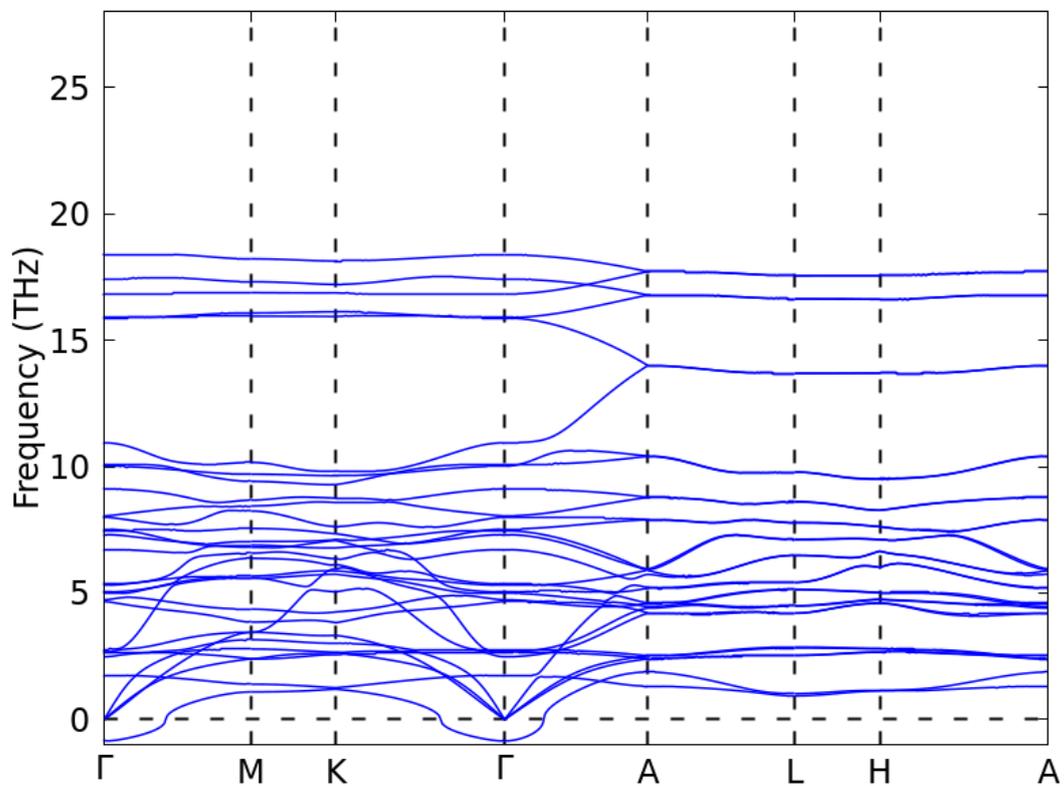

Figure S8: Phonon dispersion for AsNCa$_3$ antiperovskite in 2H Hexagonal Phase.

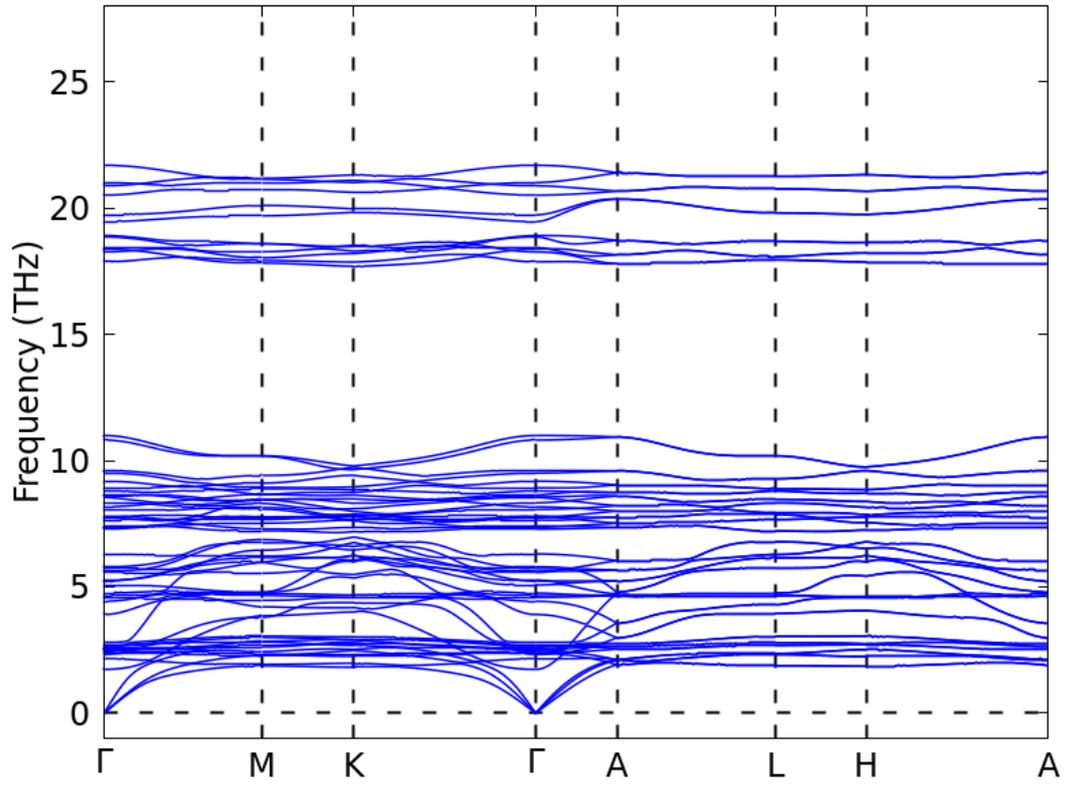

Figure S9: Phonon dispersion for AsNCa$_3$ antiperovskite in 4H Hexagonal Phase.

# S7 Thermodynamic Properties

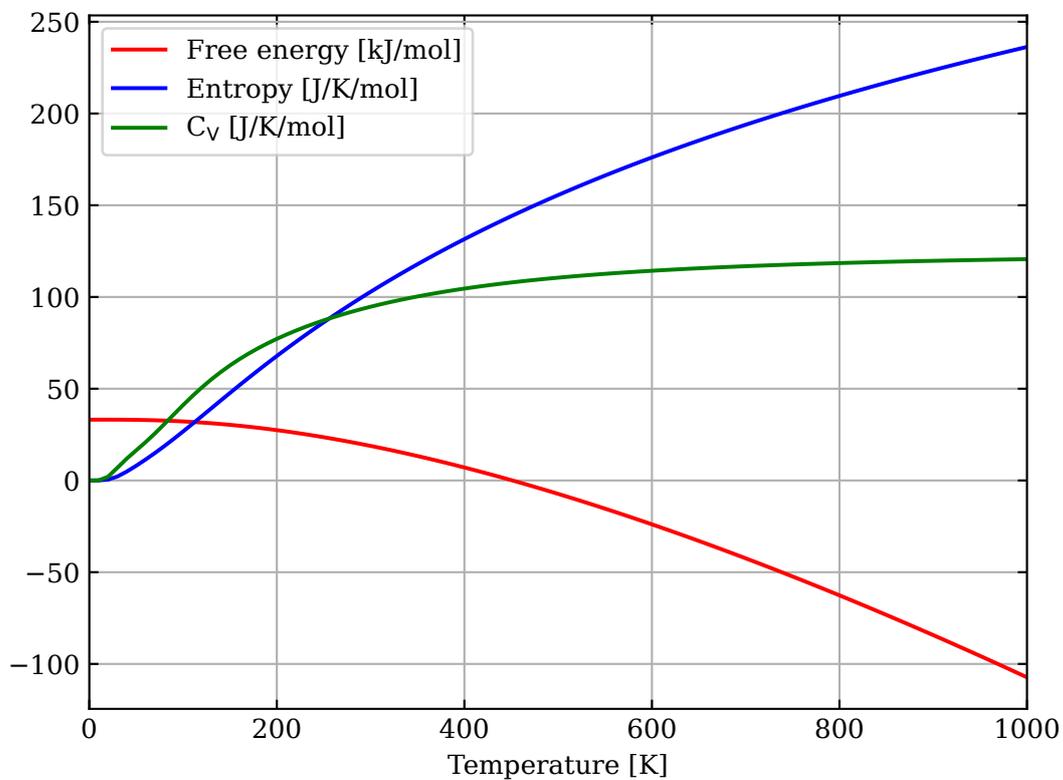

Figure S10: Thermodynamic properties for AsNCa$_3$ antiperovskite in Ideal Cubic Phase, represented by Gibbs Free Energy (red curves), Entropy (blue curves) and constant volume Heat Capacity $C_v$ (green curves).

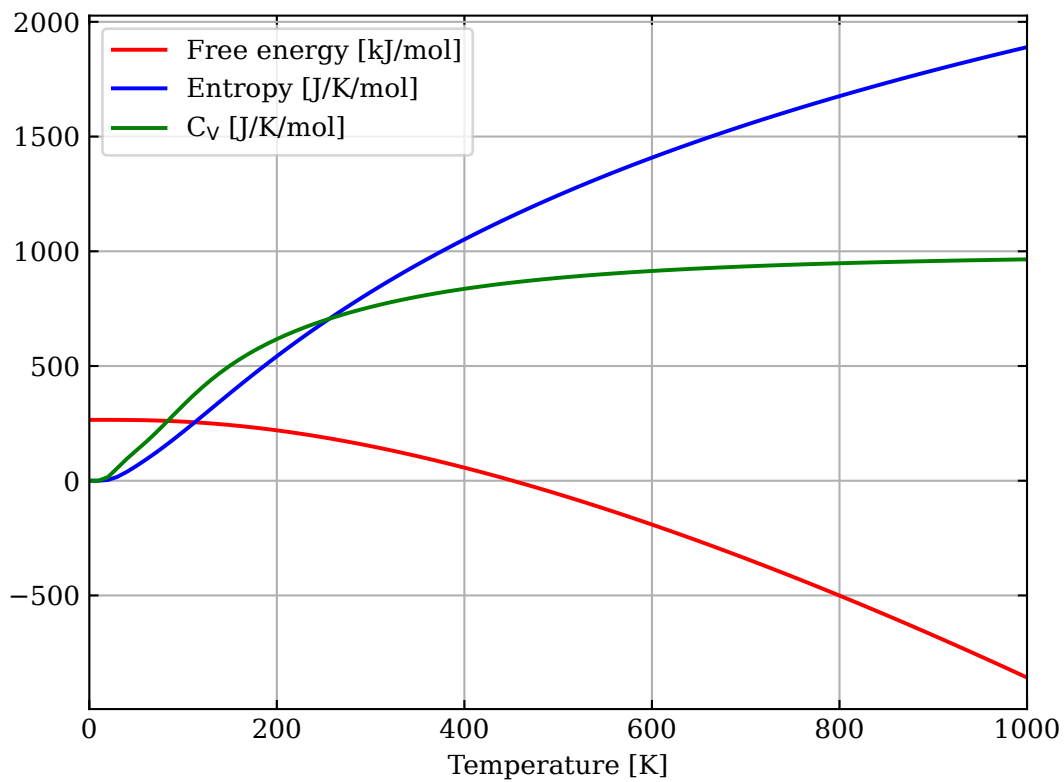

Figure S11: Thermodynamic properties for AsNCa$_3$ antiperovskite in Super Cubic Phase, represented by Gibbs Free Energy (red curves), Entropy (blue curves) and constant volume Heat Capacity $C_v$ (green curves).

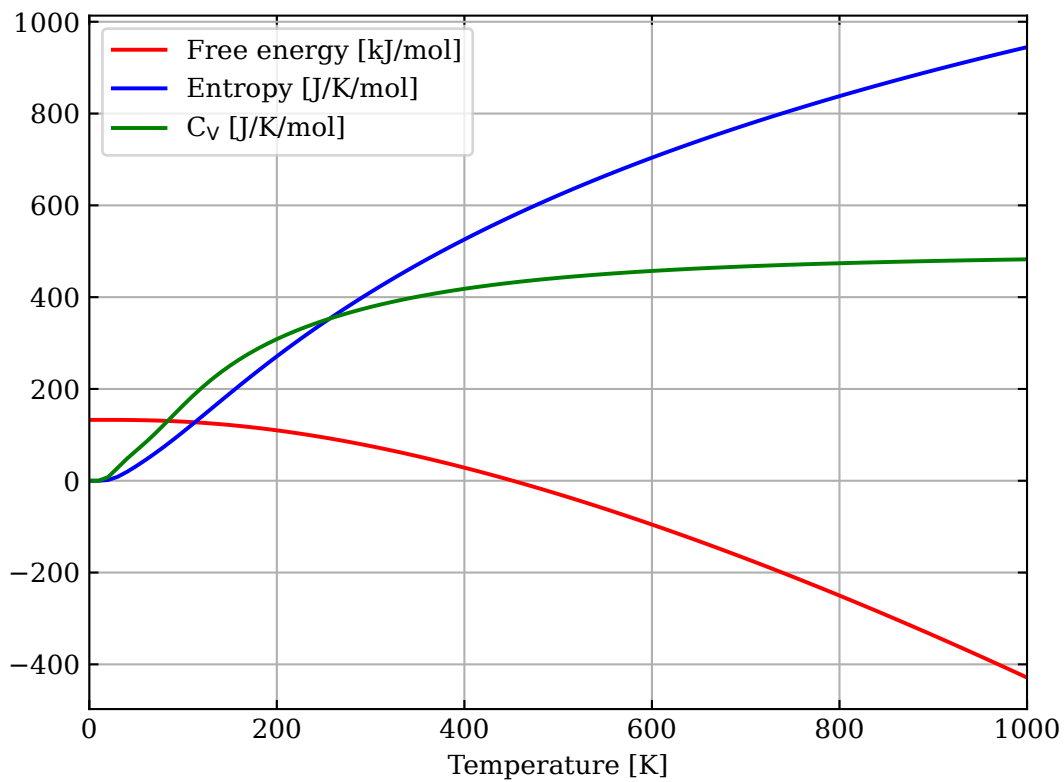

Figure S12: Thermodynamic properties for AsNCa$_3$ antiperovskite in Orthorhombic Phase, represented by Gibbs Free Energy (red curves), Entropy (blue curves) and constant volume Heat Capacity $C_v$ (green curves).

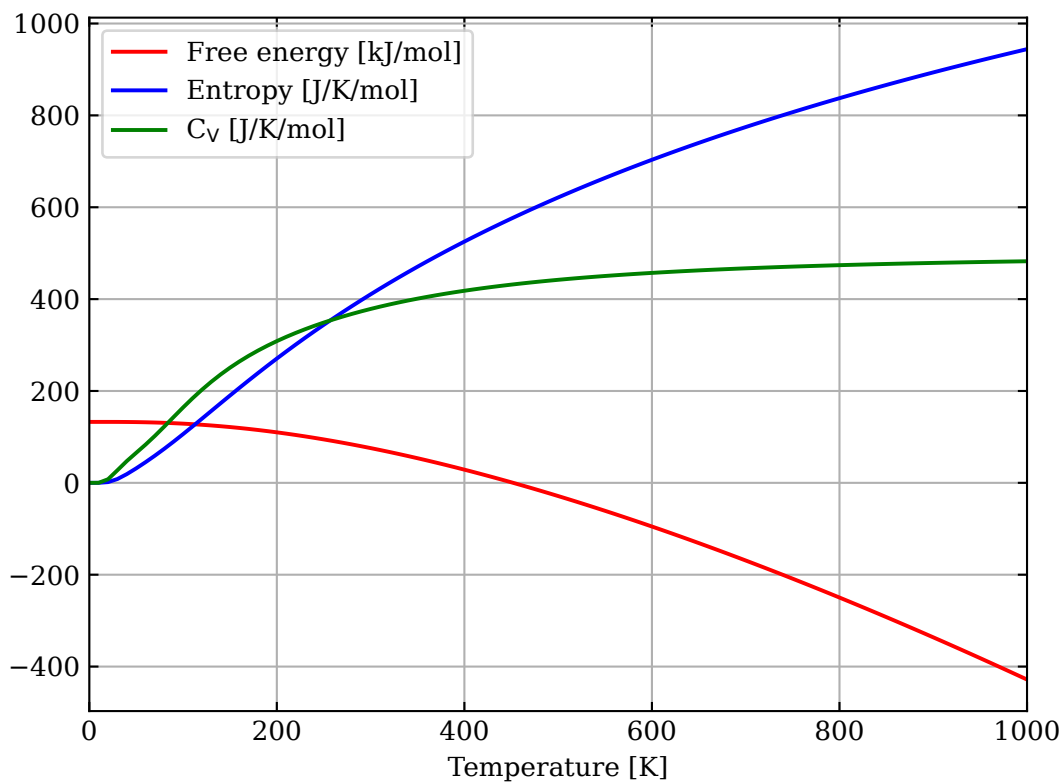

Figure S13: Thermodynamic properties for AsNCa$_3$ antiperovskite in Tetragonal Phase, represented by Gibbs Free Energy (red curves), Entropy (blue curves) and constant volume Heat Capacity $C_v$ (green curves).

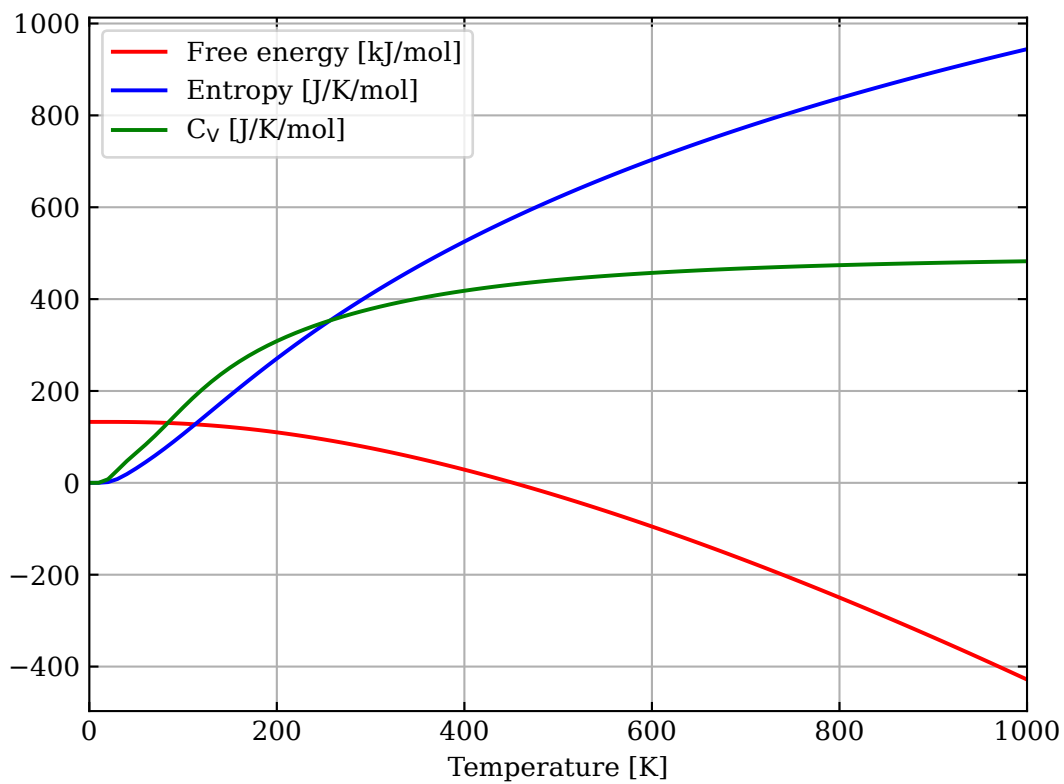

Figure S14: Thermodynamic properties for AsNCa$_3$ antiperovskite in Black Phase, represented by Gibbs Free Energy (red curves), Entropy (blue curves) and constant volume Heat Capacity $C_v$ (green curves).

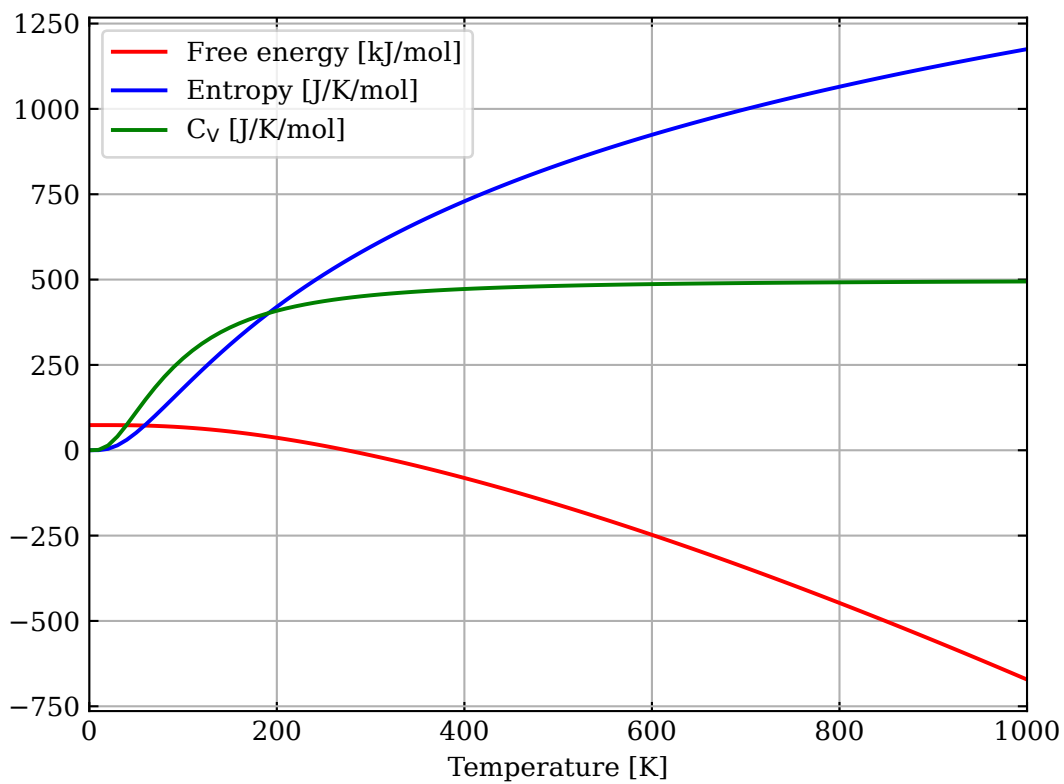

Figure S15: Thermodynamic properties for AsNCa$_3$ antiperovskite in Yellow Phase, represented by Gibbs Free Energy (red curves), Entropy (blue curves) and constant volume Heat Capacity $C_v$ (green curves).

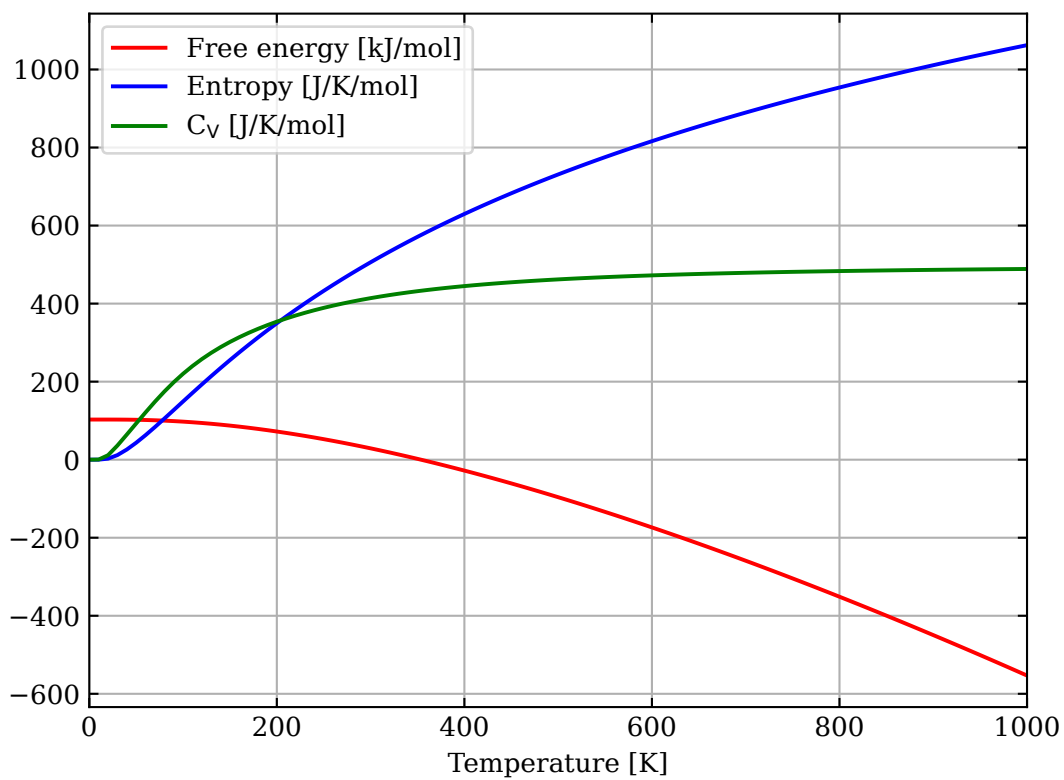

Figure S16: Thermodynamic properties for AsNCa$_3$ antiperovskite in 4H Hexagonal Phase, represented by Gibbs Free Energy (red curves), Entropy (blue curves) and constant volume Heat Capacity $C_v$ (green curves).

# S8  Density of States

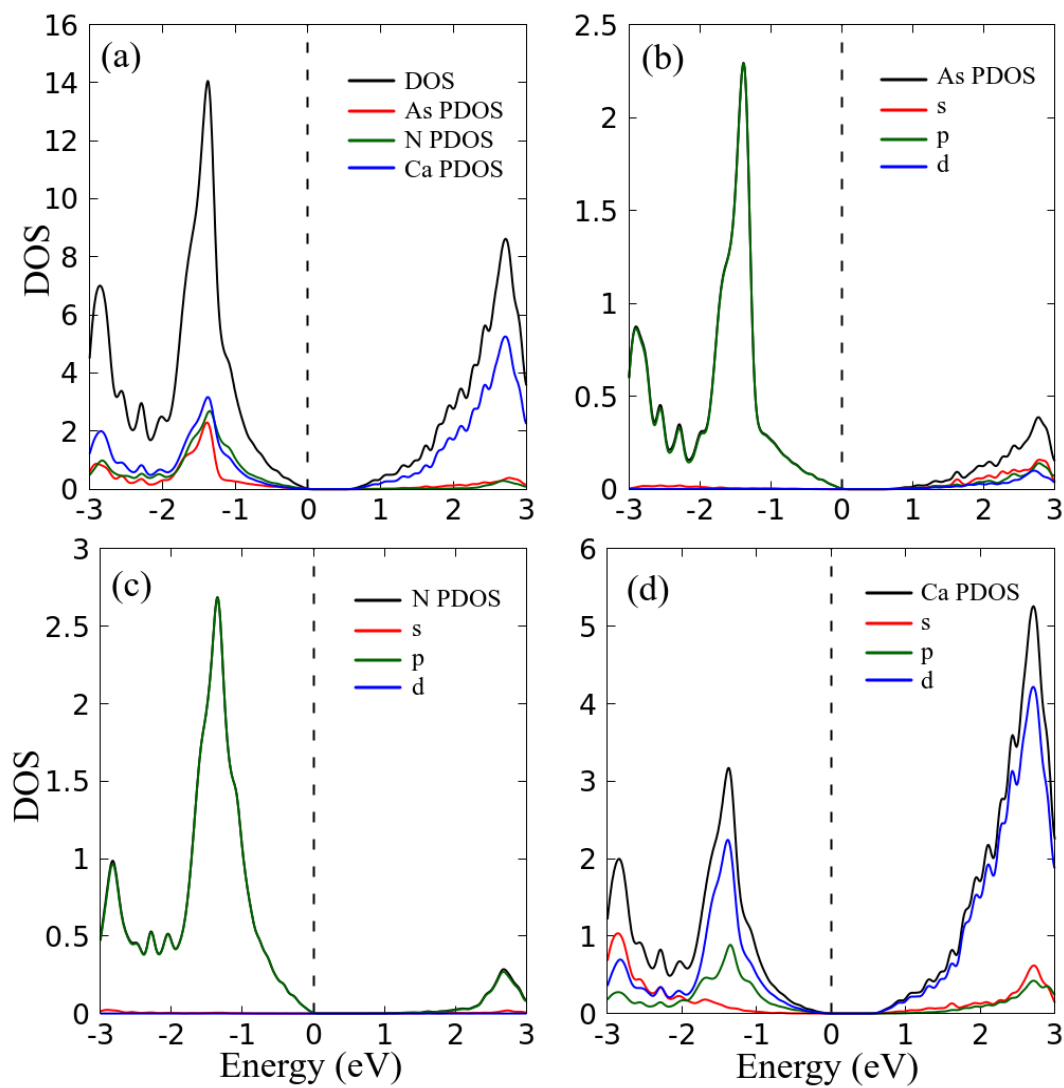

Figure S17: DOS and PDOS for AsNCa$_3$ antiperovskite in Ideal Cubic Phase.

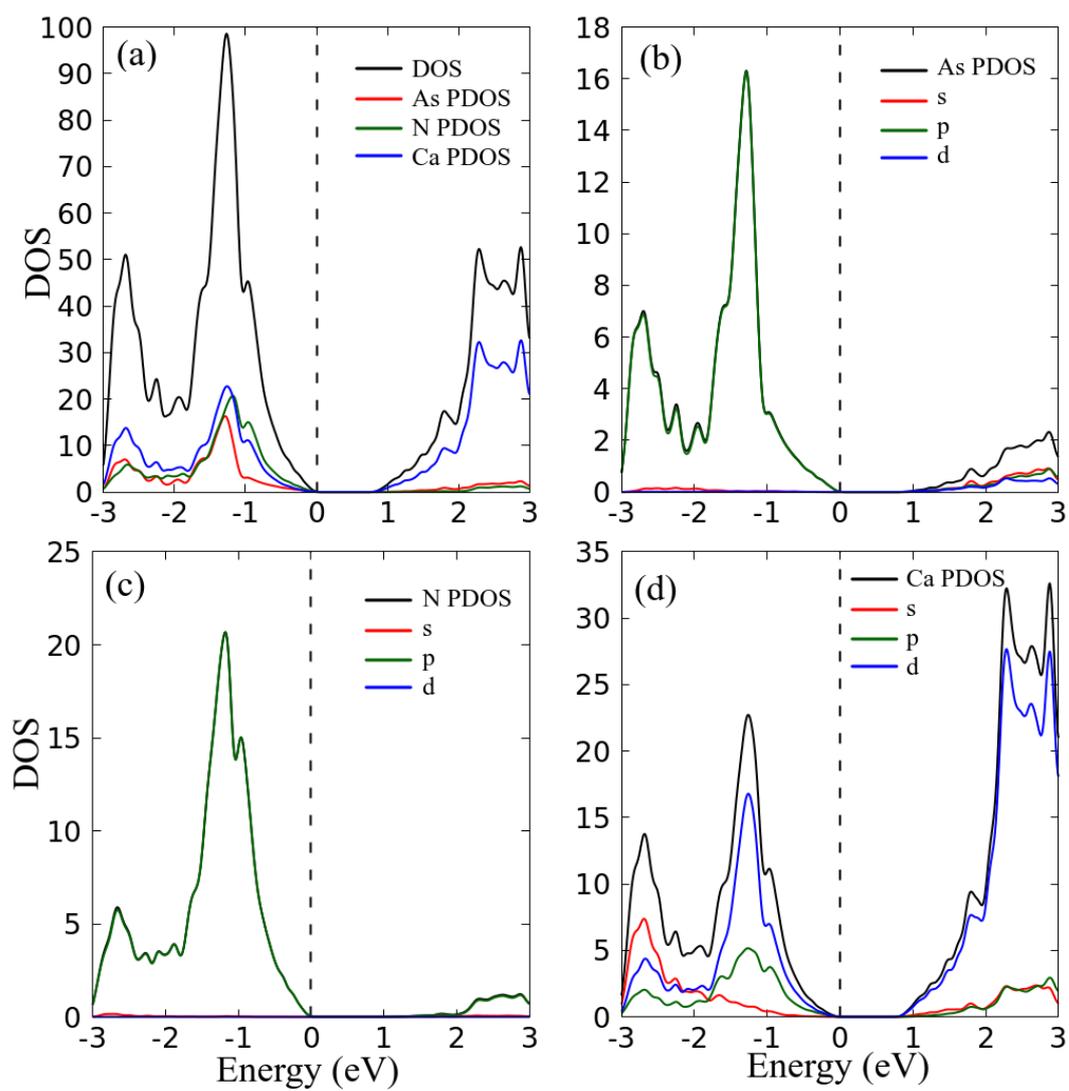

Figure S18: DOS and PDOS for AsNCa$_3$ antiperovskite in Super Cubic Phase.

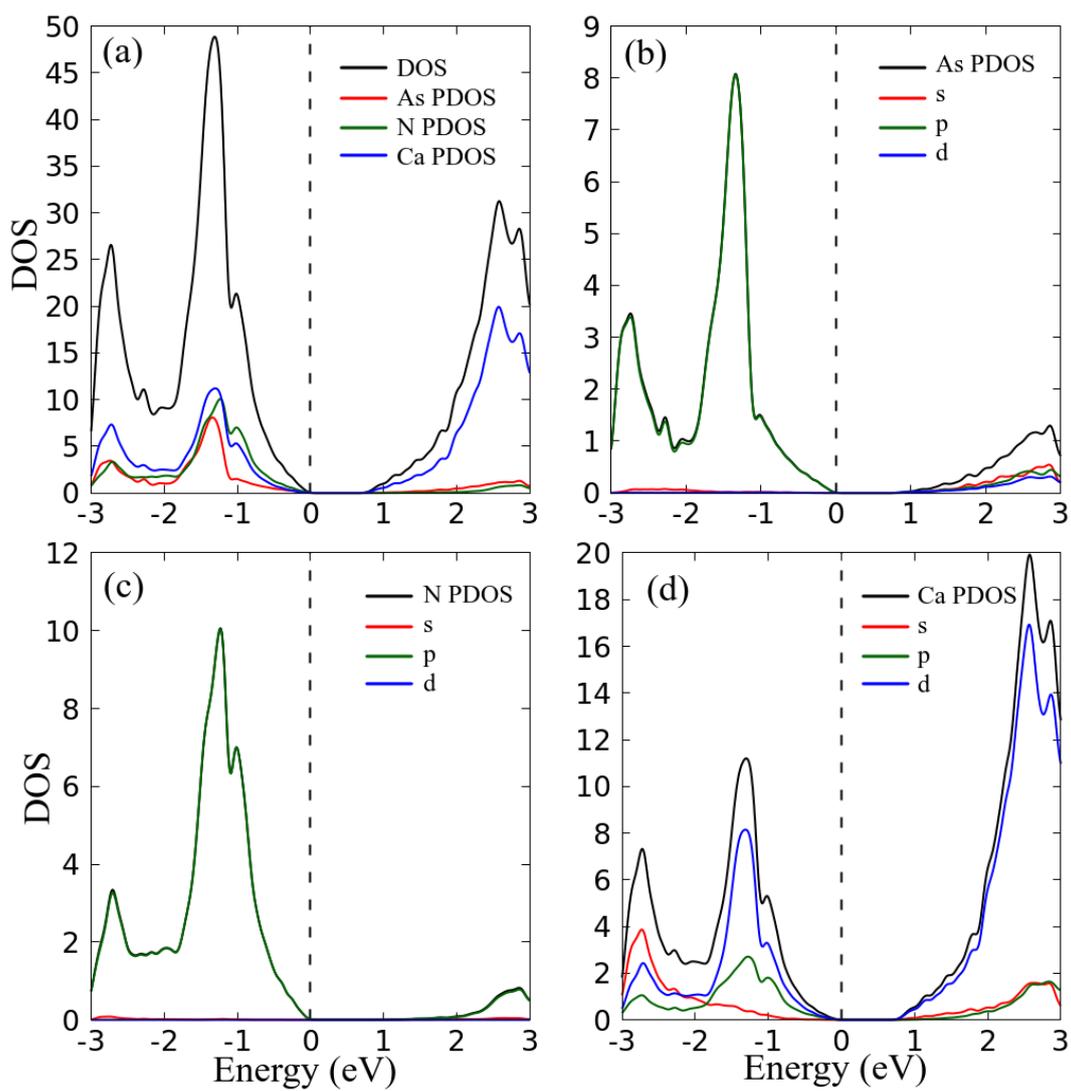

Figure S19: DOS and PDOS for AsNCa$_3$ antiperovskite in Orthorhombic Phase.

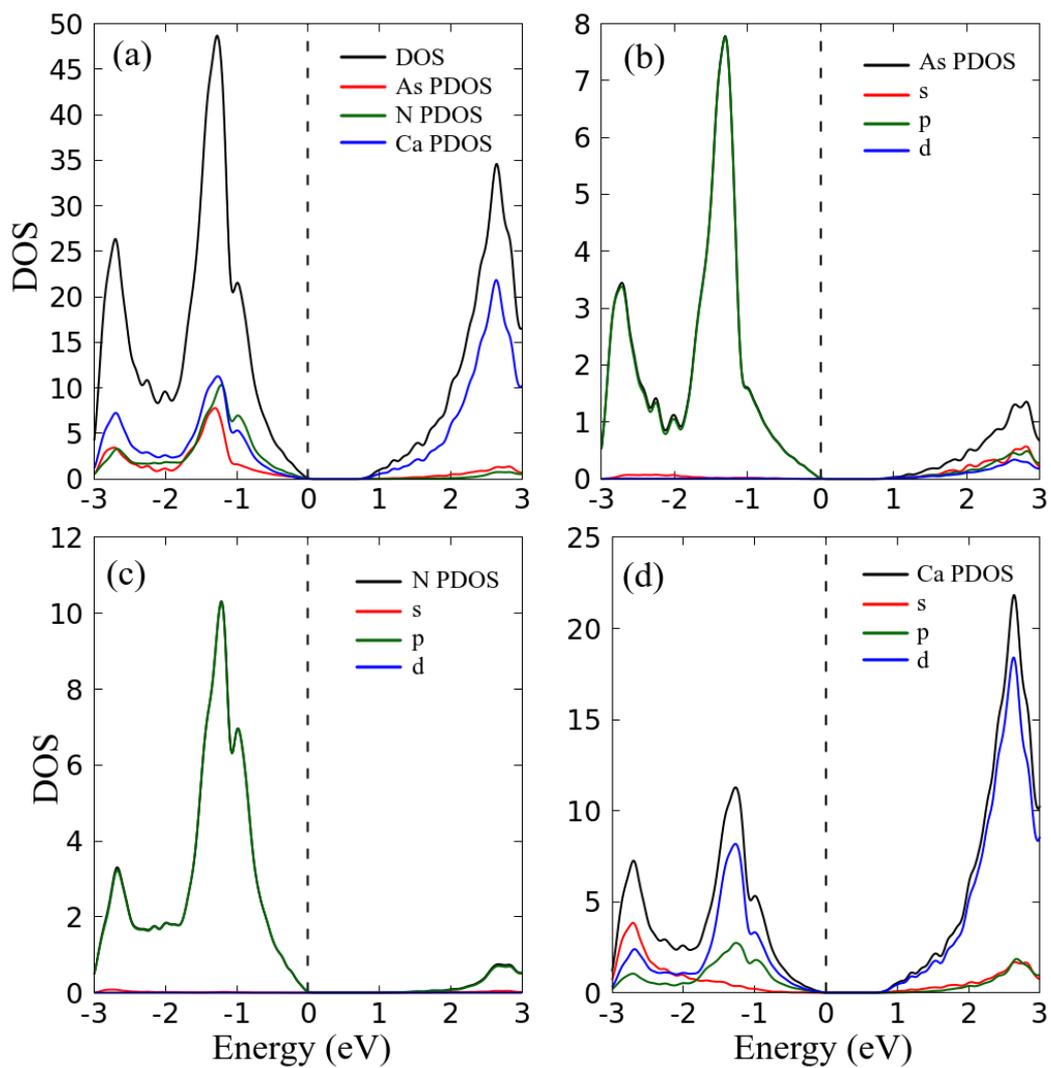

Figure S20: DOS and PDOS for AsNCa$_3$ antiperovskite in Tetragonal Phase.

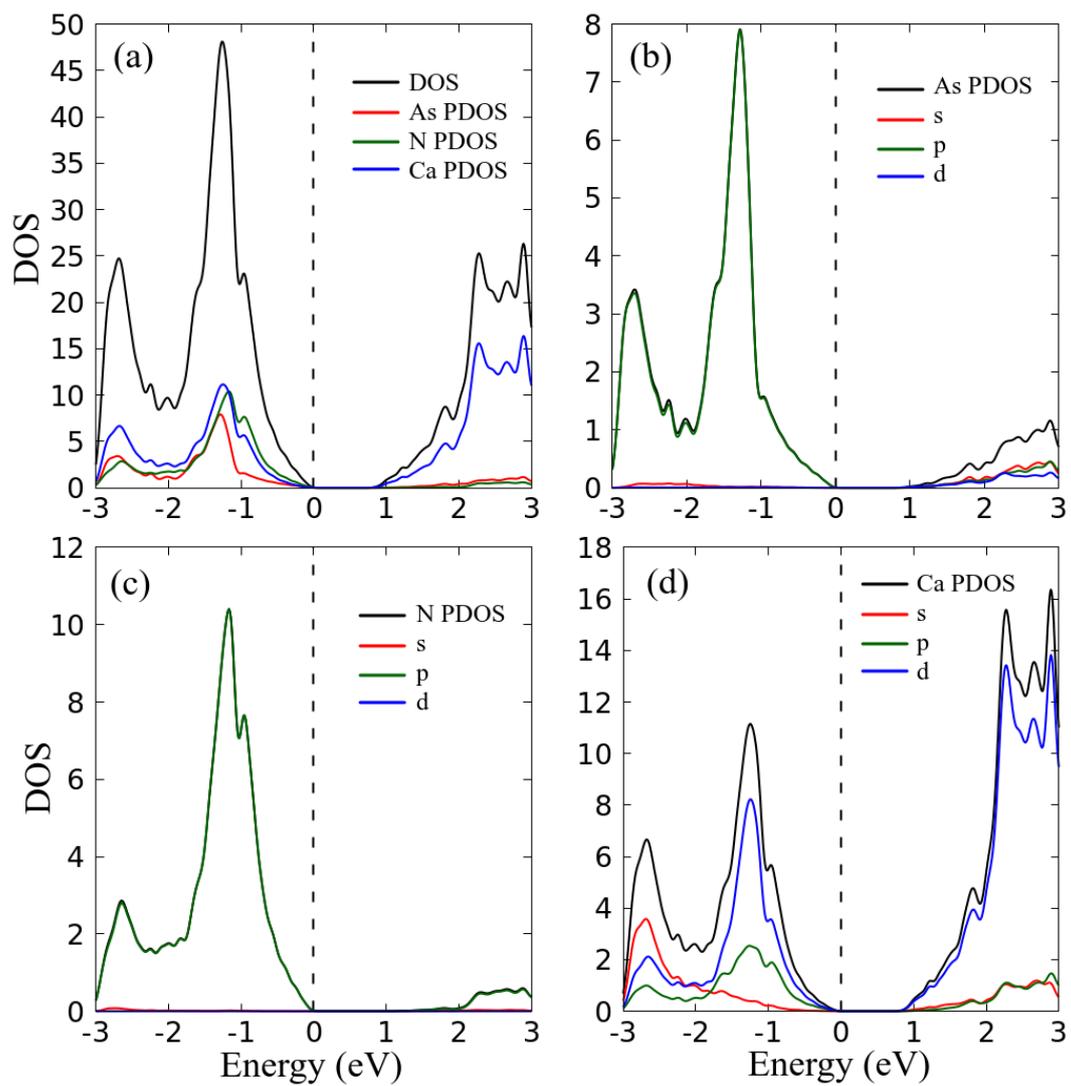

Figure S21: DOS and PDOS for AsNCa$_3$ antiperovskite in Black Phase.

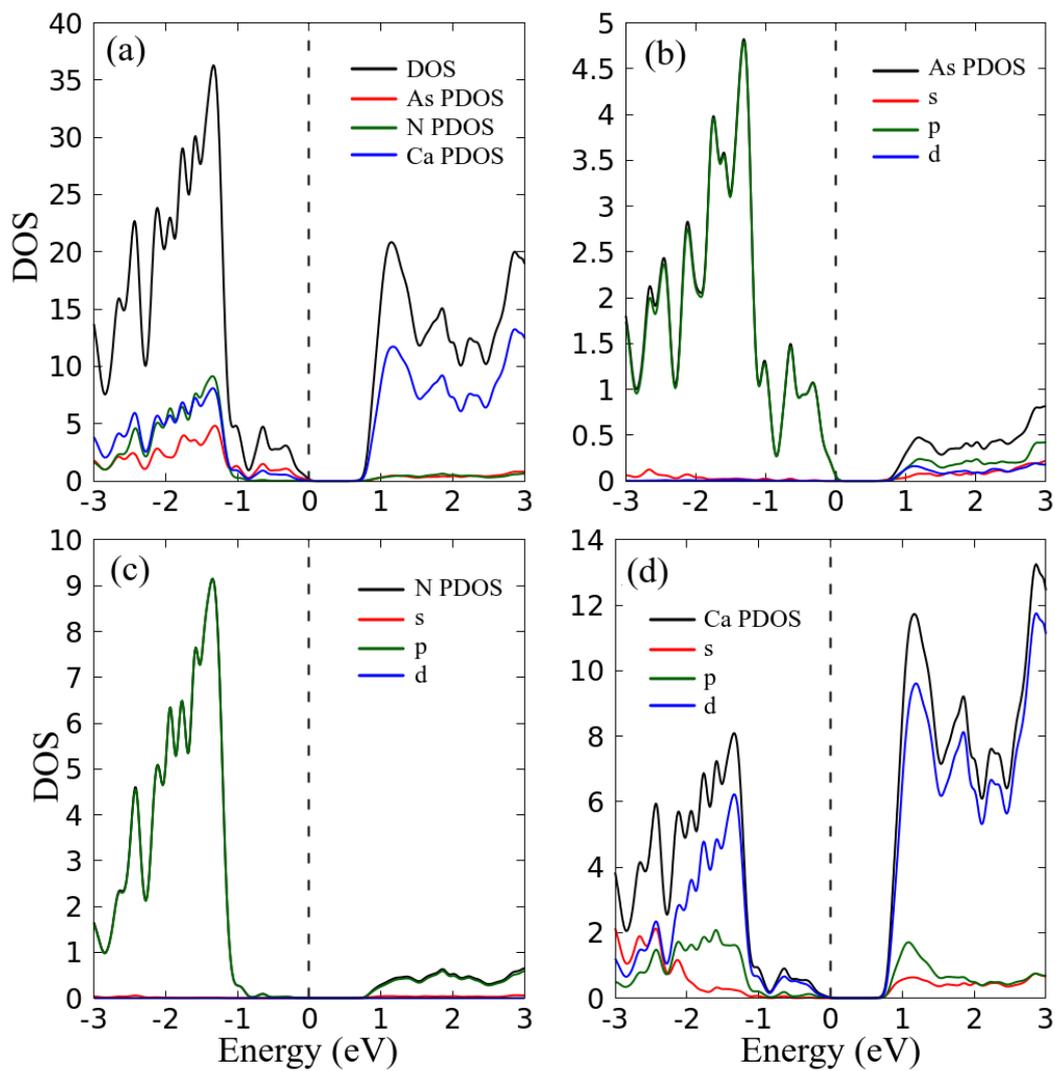

Figure S22: DOS and PDOS for AsNCa$_3$ antiperovskite in Yellow Phase.

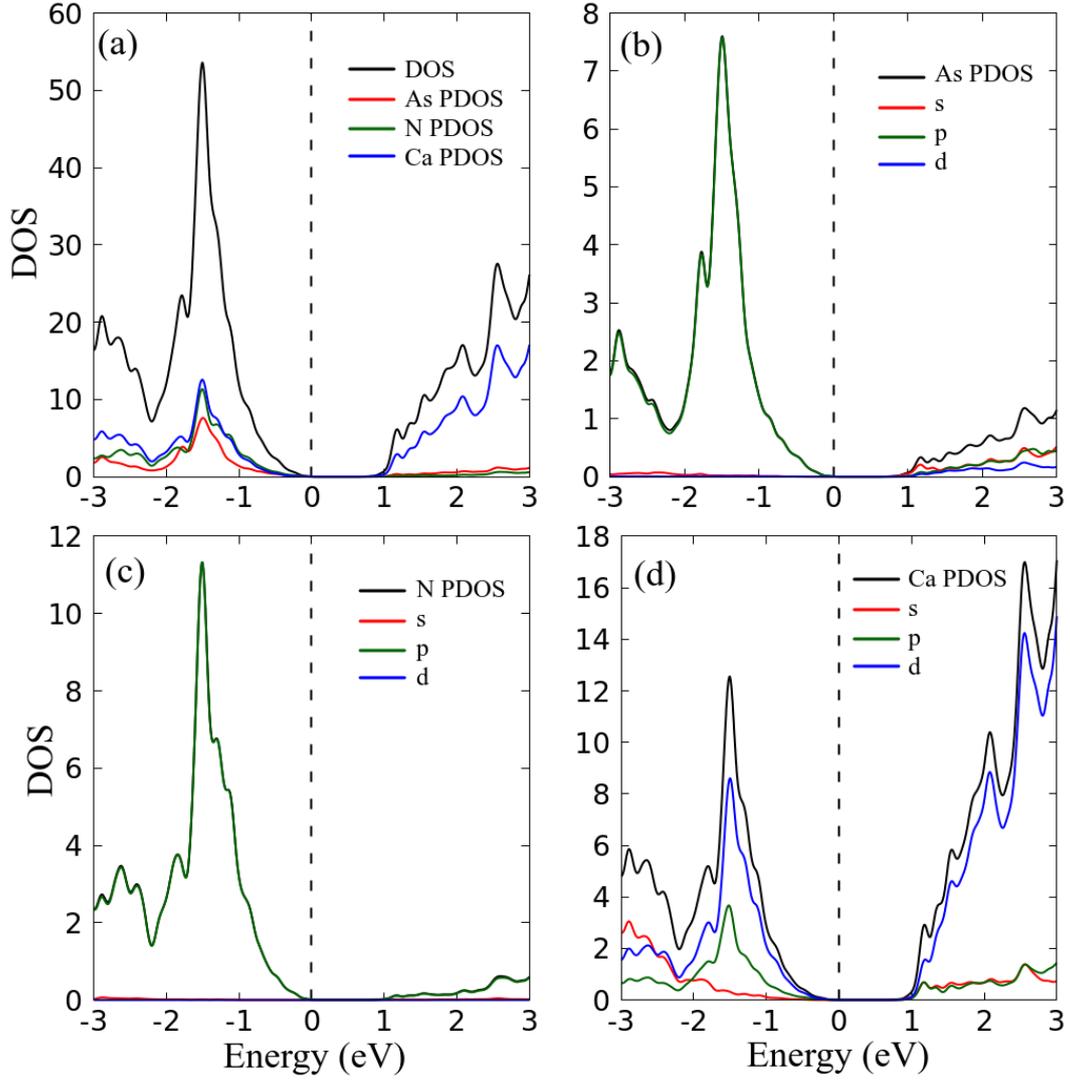

Figure S23: DOS and PDOS for AsNCa$_3$ antiperovskite in 4H Hexagonal Phase.

## S9  Electronic Band Structure

Table S4: AsNCa$_3$ PBE ($E_g^{PBE}$) and HSE06 ($E_g^{HSE06}$) electronic band gaps for each stable structural phases. All band gaps are direct.

| Phase | $E_g^{PBE}$ (eV) | $E_g^{HSE06}$ (eV) |
|---|---|---|
| Ideal Cubic | 0.69 | 1.49 |
| Super Cubic | 0.90 | 1.72 |
| Tetragonal | 0.83 | 1.65 |
| Orthorhombic | 0.83 | 1.64 |
| Black | 0.90 | 1.72 |
| Yellow | 0.67 | 1.33 |
| 4H Hexagonal | 0.77 | 1.45 |

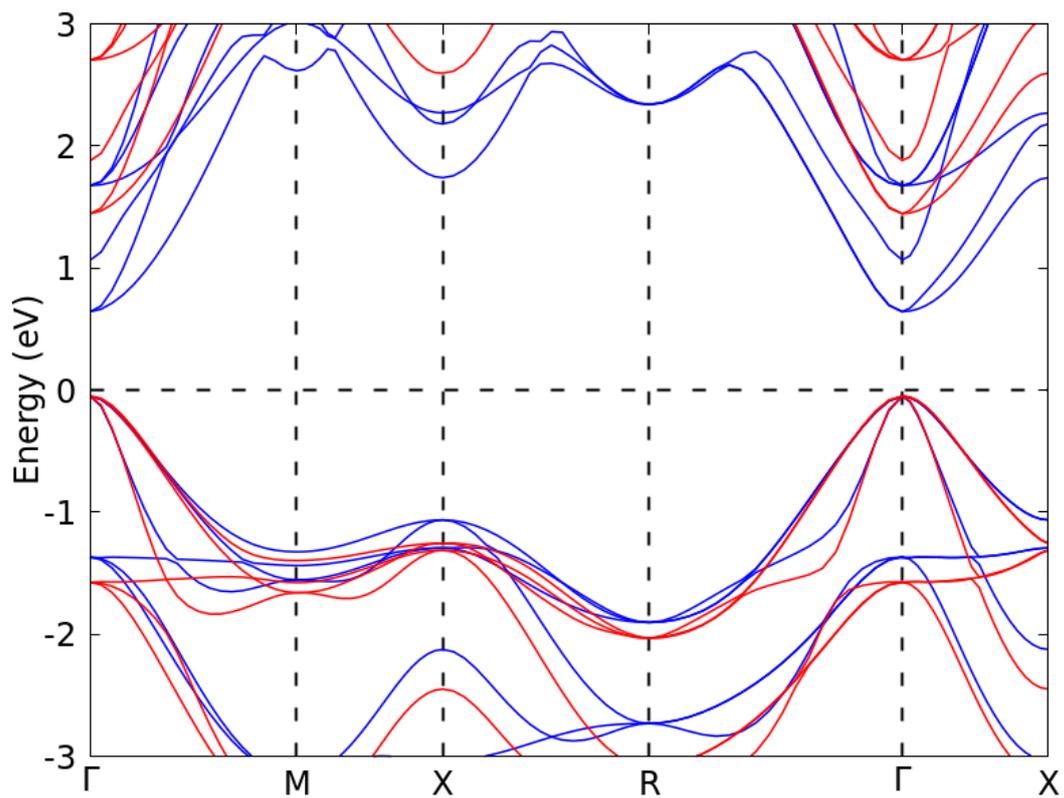

Figure S24: Band structure for AsNCa$_3$ antiperovskite in Ideal Cubic Phase with PBE (blue curves) and HSE06 (red curves) functionals.

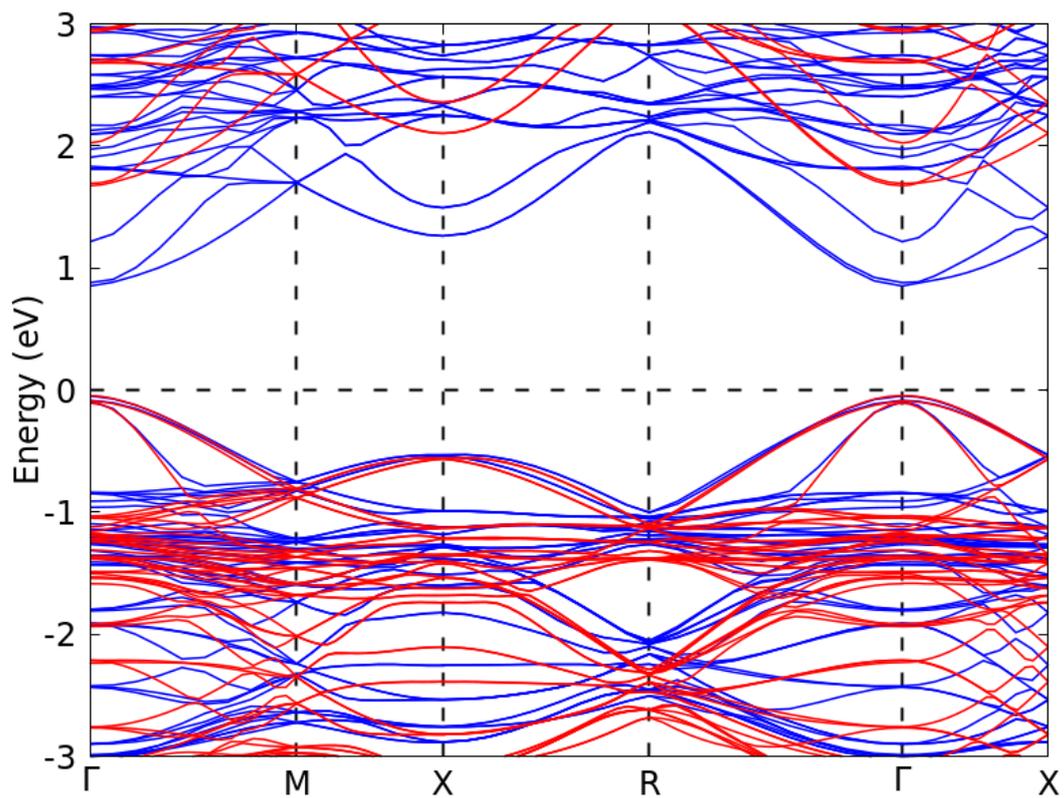

Figure S25: Band structure for AsNCa$_3$ antiperovskite in Super Cubic Phase with PBE (blue curves) and HSE06 (red curves) functionals.

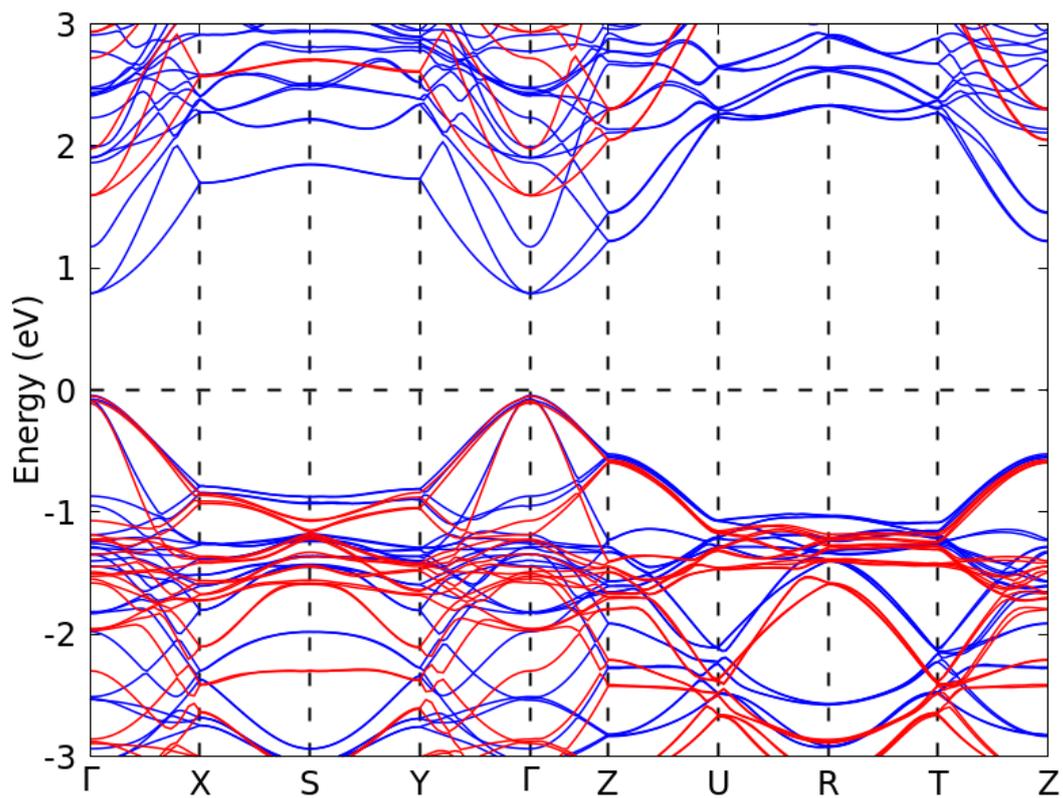

Figure S26: Band structure for AsNCa$_3$ antiperovskite in Orthorhombic Phase with PBE (blue curves) and HSE06 (red curves) functionals.

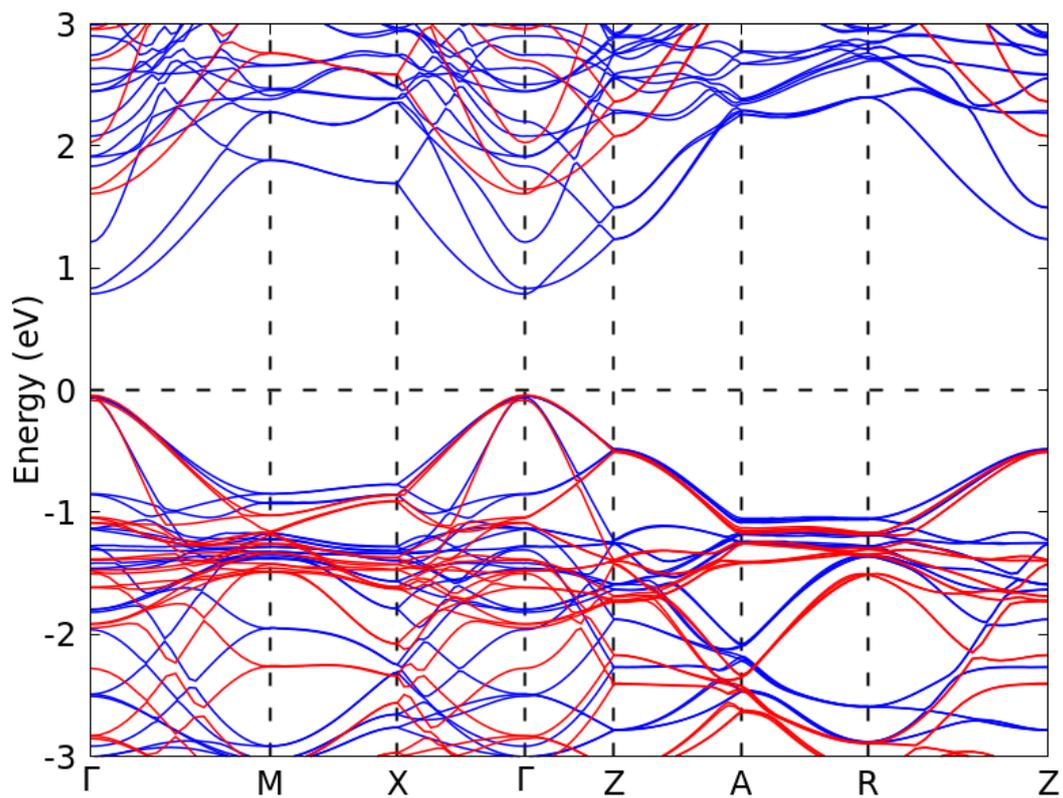

Figure S27: Band structure for AsNCa$_3$ antiperovskite in Tetragonal Phase with PBE (blue curves) and HSE06 (red curves) functionals.

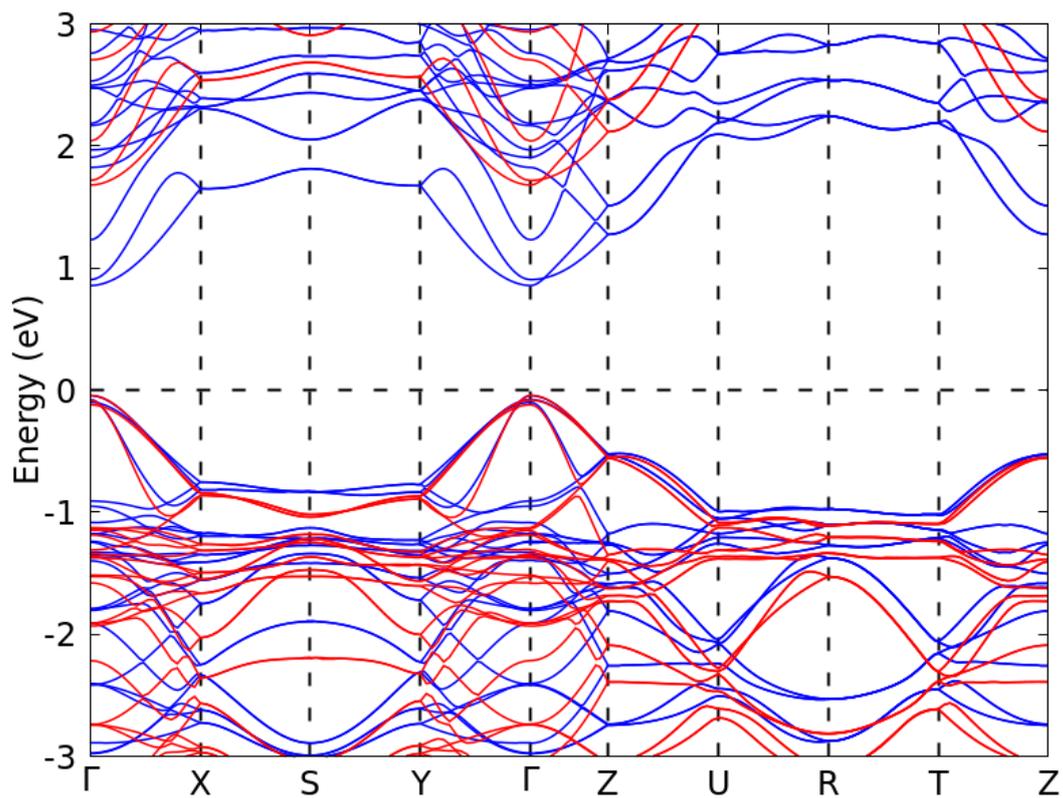

Figure S28: Band structure for $AsNCa_3$ antiperovskite in Black Phase with PBE (blue curves) and HSE06 (red curves) functionals.

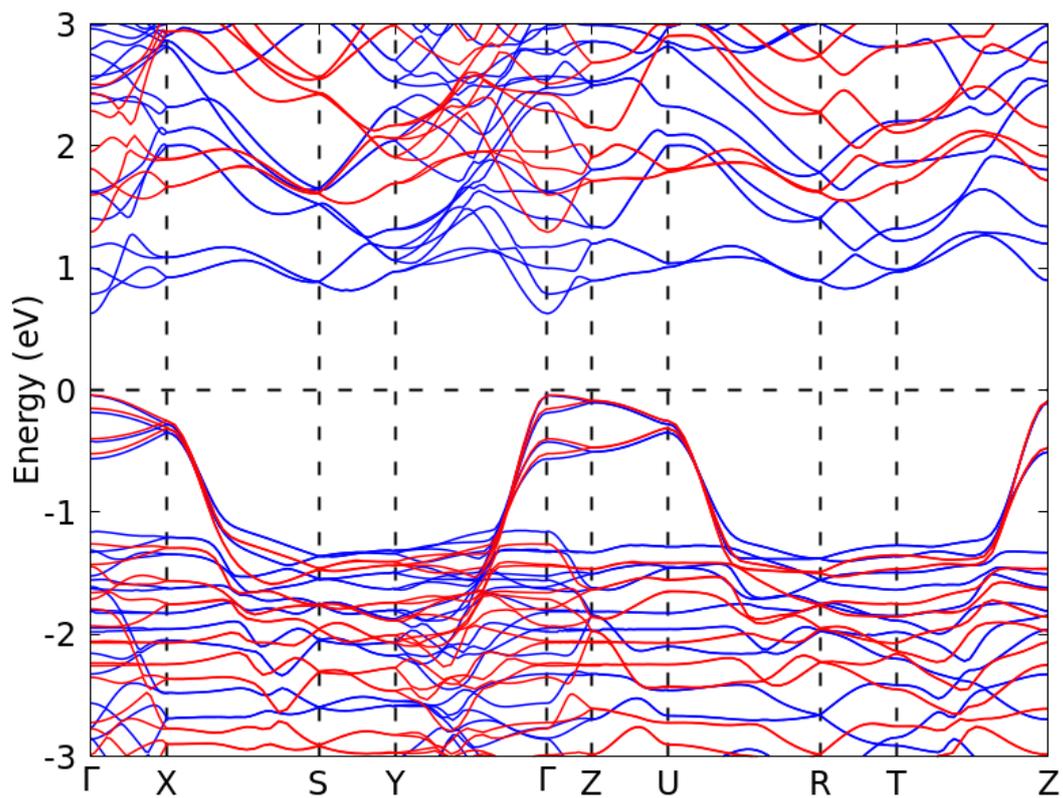

Figure S29: Band structure for AsNCa$_3$ antiperovskite in Yellow Phase with PBE (blue curves) and HSE06 (red curves) functionals.

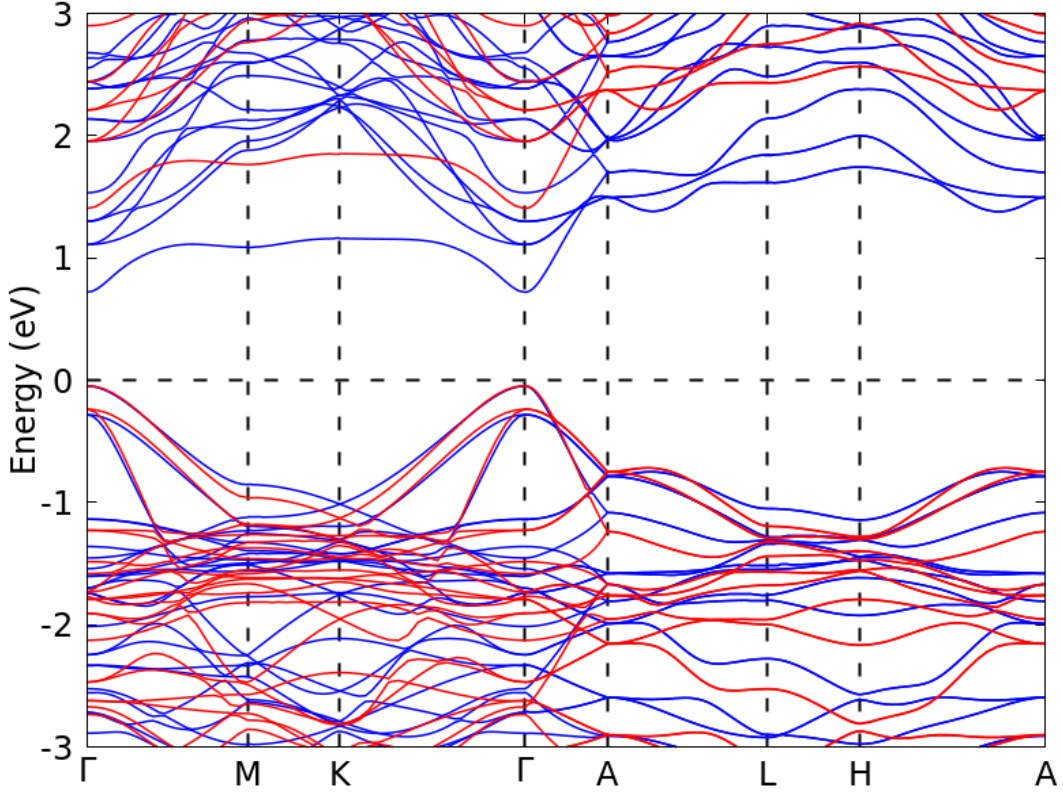

Figure S30: Band structure for AsNCa$_3$ antiperovskite in 4H Hexagonal Phase with PBE (blue curves) and HSE06 (red curves) functionals.

## S10 Optical Properties

Table S5: Parameters used for Optical properties simulations: **k**-points density, $R_k$ (Å$^{-1}$) and their correspondent **k**-mesh, $n_v$, number of valence bands, $n_c$, number of conduction bands and dielectric function smearing $\eta$ (eV).

| Phase | $R_k$ | **k**-mesh | $n_c$ | $n_v$ | $\eta$ |
|---|---|---|---|---|---|
| Ideal Cubic | 120 | $25 \times 25 \times 25$ | 6 | 6 | 0.05 |
| Super Cubic | 120 | $13 \times 13 \times 13$ | 23 | 36 | 0.05 |
| Tetragonal | 120 | $18 \times 18 \times 13$ | 14 | 18 | 0.05 |
| Orthorhombic | 120 | $18 \times 18 \times 13$ | 14 | 18 | 0.05 |
| Black | 120 | $18 \times 18 \times 13$ | 13 | 18 | 0.05 |
| Yellow | 120 | $15 \times 30 \times 9$ | 14 | 16 | 0.05 |
| 4H Hexagonal | 120 | $21 \times 21 \times 11$ | 16 | 17 | 0.05 |

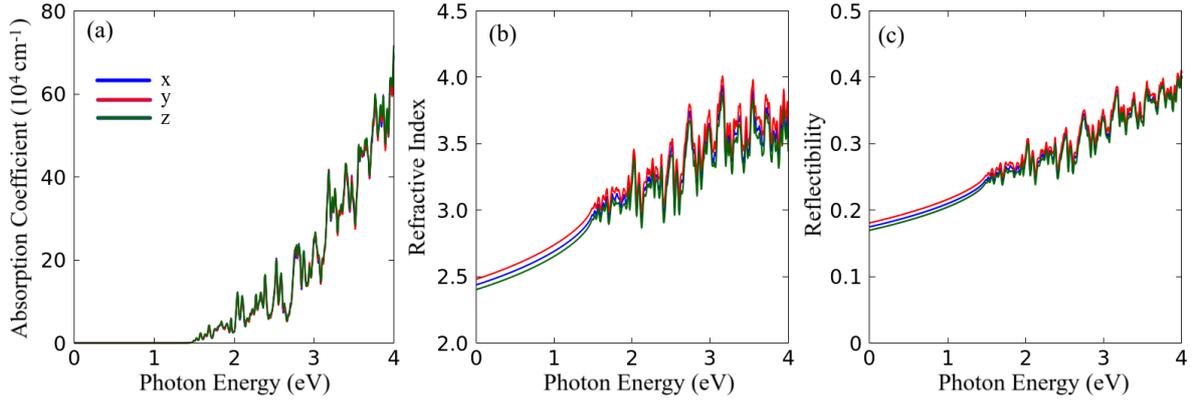

Figure S31: Optical properties for AsNCa$_3$ antiperovskite in Ideal Cubic Phase: (a) Absorption Coefficient, (b) Refractive Index and (c) Reflectibility, under linear light polarization at $\hat{x}$ (blue curves), $\hat{y}$ (red curves) and $\hat{z}$ (green curves) directions.

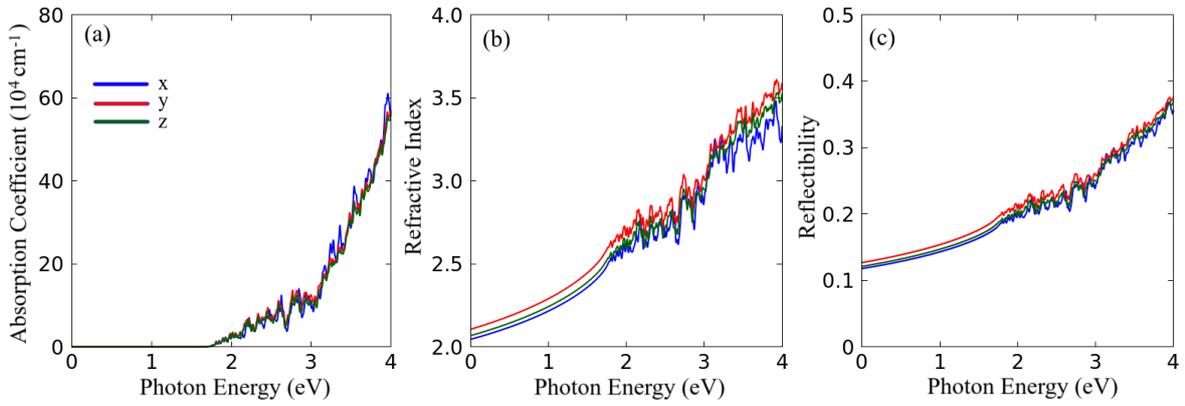

Figure S32: Optical properties for AsNCa$_3$ antiperovskite in Super Cubic Phase: (a) Absorption Coefficient, (b) Refractive Index and (c) Reflectibility, under linear light polarization at $\hat{x}$ (blue curves), $\hat{y}$ (red curves) and $\hat{z}$ (green curves) directions.

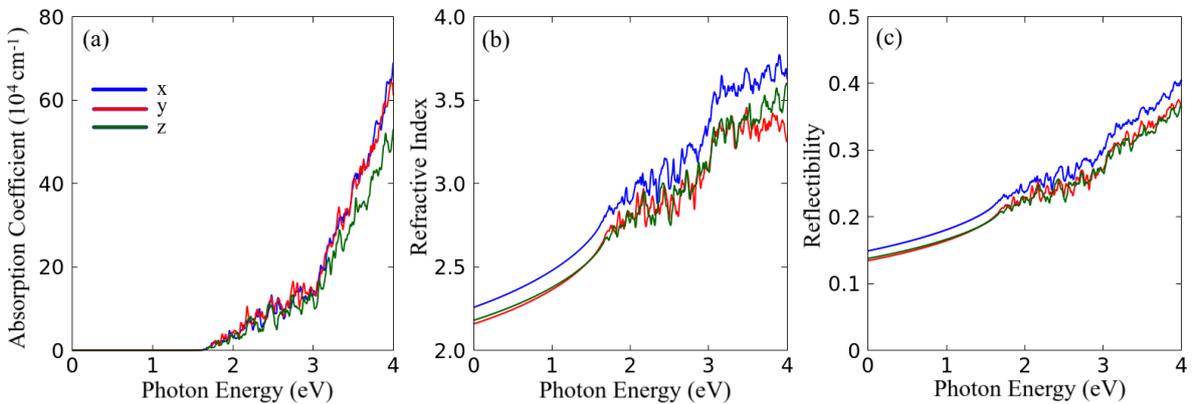

Figure S33: Optical properties for AsNCa$_3$ antiperovskite in Orthorhombic Phase: (a) Absorption Coefficient, (b) Refractive Index and (c) Reflectibility, under linear light polarization at $\hat{x}$ (blue curves), $\hat{y}$ (red curves) and $\hat{z}$ (green curves) directions.

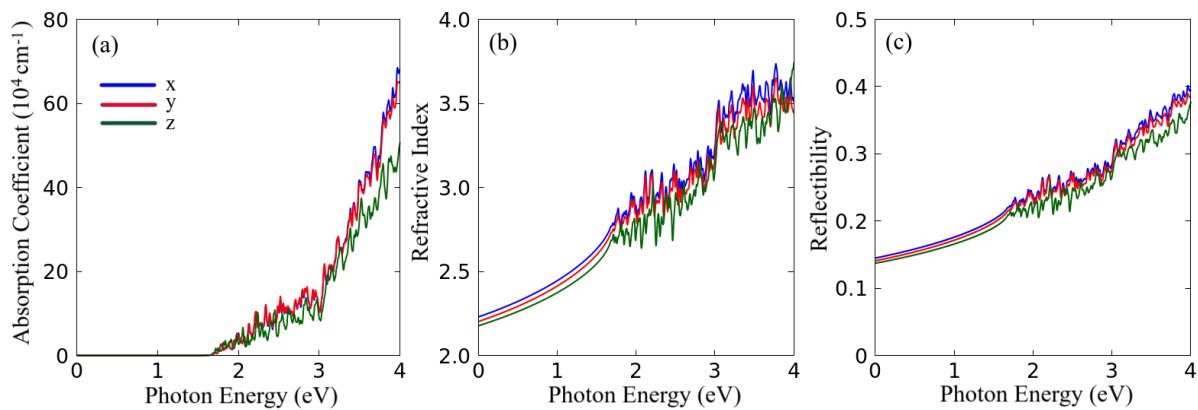

Figure S34: Optical properties for AsNCa$_3$ antiperovskite in Tetragonal Phase: (a) Absorption Coefficient, (b) Refractive Index and (c) Reflectibility, under linear light polarization at $\hat{x}$ (blue curves), $\hat{y}$ (red curves) and $\hat{z}$ (green curves) directions.

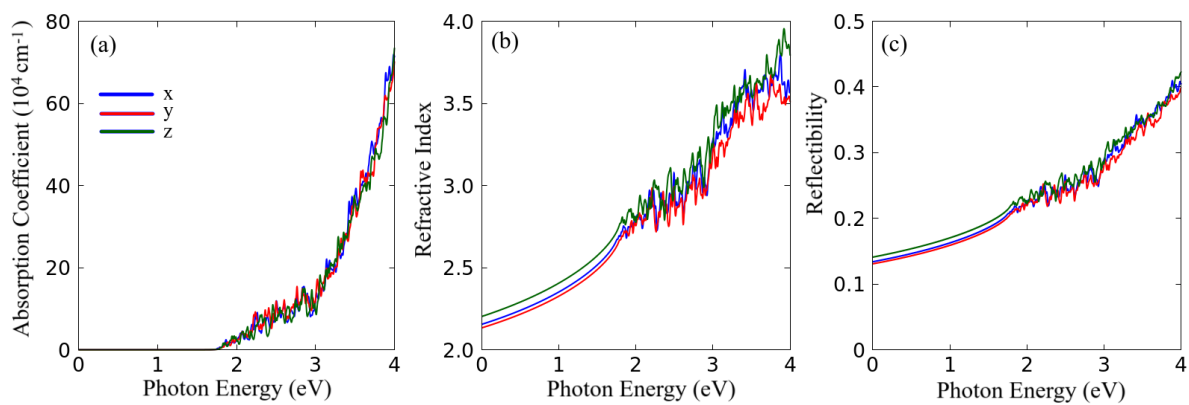

Figure S35: Optical properties for AsNCa$_3$ antiperovskite in Black Phase: (a) Absorption Coefficient, (b) Refractive Index and (c) Reflectibility, under linear light polarization at $\hat{x}$ (blue curves), $\hat{y}$ (red curves) and $\hat{z}$ (green curves) directions.

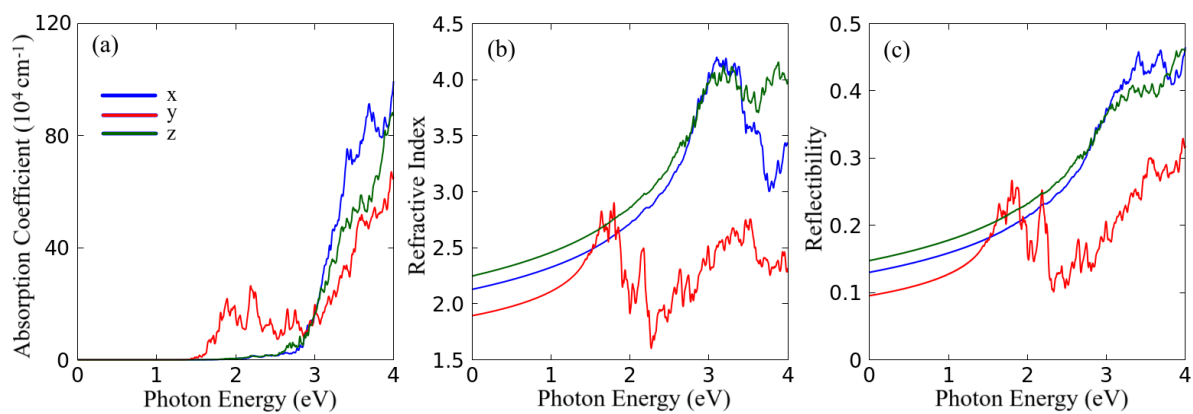

Figure S36: Optical properties for AsNCa$_3$ antiperovskite in Yellow Phase: (a) Absorption Coefficient, (b) Refractive Index and (c) Reflectibility, under linear light polarization at $\hat{x}$ (blue curves), $\hat{y}$ (red curves) and $\hat{z}$ (green curves) directions.

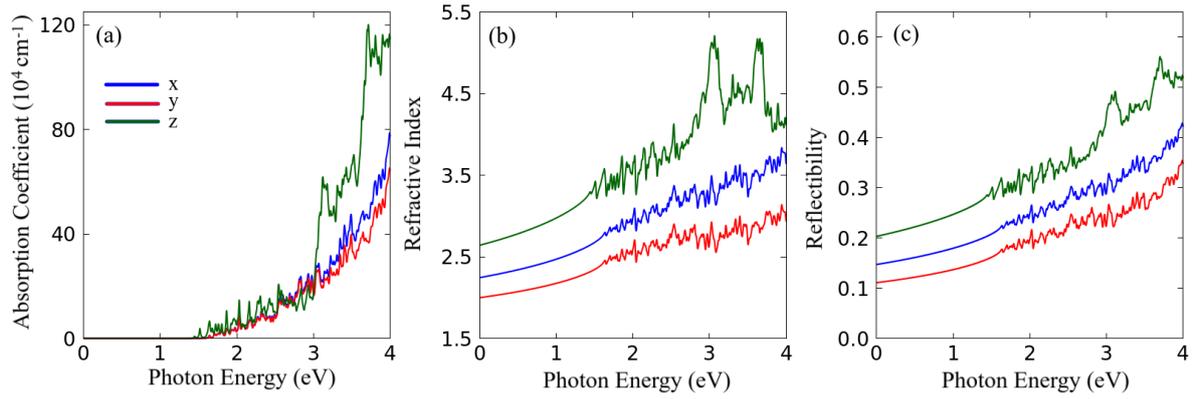

Figure S37: Optical properties for AsNCa$_3$ antiperovskite in 4H Hexagonal Phase: (a) Absorption Coefficient, (b) Refractive Index and (c) Reflectibility, under linear light polarization at $\hat{x}$ (blue curves), $\hat{y}$ (red curves) and $\hat{z}$ (green curves) directions.

## S11  Insights in Solar Harvesting Efficiency

Table S6: Maximum achieved PCE at the IPA level, short circuit current density $J_{sc}$ (W/V.m$^2$), open circuit voltage $V_{oc}$ (V), recombination factor $fr$, fill factor $FF$ (%), material thickness $\Delta$ (µm), power conversion efficiency obtained by SLME (PCE$^{SLME}$) (%), all calculated at T = 300K.

| Phase | $J_{sc}$ | $V_{oc}$ | $fr$ | $FF$ | $\Delta$ | PCE$^{SLME}$ |
|---|---|---|---|---|---|---|
| Ideal Cubic | 289.96 | 1.20 | 1.00 | 89.76 | 1.00 | 31.23 |
| Super Cubic | 212.66 | 1.42 | 1.00 | 91.03 | 1.00 | 27.50 |
| Tetragonal | 235.24 | 1.35 | 1.00 | 90.67 | 1.00 | 28.84 |
| Orthorhombic | 238.95 | 1.34 | 1.00 | 90.61 | 1.00 | 29.05 |
| Black | 310.62 | 1.09 | 1.00 | 89.00 | 1.00 | 30.24 |
| Yellow | 209.57 | 1.43 | 1.00 | 91.06 | 1.00 | 27.25 |
| 4H Hexagonal | 294.84 | 1.16 | 1.00 | 89.52 | 1.00 | 30.71 |